\documentclass[11pt,twoside,english]{revtex4-1}
\pdfoutput=1
\usepackage[T1]{fontenc}
\usepackage[latin9]{inputenc}
\usepackage[a4paper]{geometry}
\usepackage{color}
\usepackage{babel}
\usepackage{varioref}
\usepackage{prettyref}
\usepackage{float}
\usepackage{amsmath}
\usepackage{amssymb}
\usepackage{graphicx}
\usepackage[unicode=true,
 bookmarks=true,bookmarksnumbered=true,bookmarksopen=false,
 breaklinks=true,pdfborder={0 0 1},backref=false,colorlinks=true]
 {hyperref}
\hypersetup{
 colorlinks=true,linktoc=all,linkcolor=blue, citecolor=blue, urlcolor=blue, filecolor=blue, pdfpagelayout=OneColumn, pdfnewwindow=true, pdfstartview=XYZ, plainpages=false}
\usepackage{dcolumn}
\newcolumntype{.}{D{.}{.}{5.3}}
\makeatletter

\providecommand{\tabularnewline}{\\}

\makeatother

\begin{document}

\title{Practical optimization of Steiner Trees via the cavity method}

\author{Alfredo Braunstein}

\email{alfredo.braunstein@polito.it}

\affiliation{DISAT, Politecnico di Torino, Corso Duca Degli Abruzzi 24, Torino,
Italy}

\affiliation{Human Genetics Foundation, Via Nizza 52, Torino, Italy}

\affiliation{Collegio Carlo Alberto, Via Real Collegio 30, Moncalieri, Italy}

\author{Anna Muntoni}

\email{anna.muntoni@polito.it}

\affiliation{DISAT, Politecnico di Torino, Corso Duca Degli Abruzzi 24, Torino,
Italy}

\begin{abstract}
The optimization version of the cavity method for single instances, called Max-Sum, has been applied in the past to the Minimum Steiner Tree Problem on Graphs and variants. Max-Sum has been shown experimentally to give asymptotically optimal results on certain types of weighted random graphs, and to give good solutions in short computation times for some types of real networks. However, the hypotheses behind the formulation and the cavity method itself limit substantially the class of instances on which the approach gives good results (or even converges). Moreover, in the standard model formulation, the diameter of the tree solution is limited by a predefined bound, that affects both computation time and convergence properties. In this work we describe two main enhancements to the Max-Sum equations to be able to cope with optimization of real-world instances. First, we develop an alternative ``\emph{flat}'' model formulation, that allows to reduce substantially the relevant configuration space, making the approach feasible on instances with large solution diameter, in particular when the number of terminal nodes is small. Second, we propose an integration between Max-Sum and three greedy heuristics. This integration allows to transform Max-Sum into a highly competitive self-contained algorithm, in which a feasible solution is given at each step of the iterative procedure. Part of this development participated on the 2014 DIMACS challenge on Steiner Problems, and we report the results here. The performance on the challenge of the proposed approach was highly satisfactory: it maintained a small gap to the best bound in most cases, and obtained best results on several instances on two different categories. We also present several improvements with respect to the version of the algorithm that participated in the competition, including new best solutions for some of the instances of the challenge.
\end{abstract}
\maketitle
\tableofcontents{}

\section{Introduction}

The cavity method has been developed for the study of disordered systems
in statistical physics and has been employed in recent years for the
design of a family of algorithmic techniques for combinatorial optimization.
Among these, it has been shown that series of network optimization
problems including the Minimum Steiner Tree (MStT) problem and variants,
e.g. the Prize-Collecting Steiner Problem on Graphs (PCSPG), can successfully
be described by a model with local constraints and solved (at least
on certain families of instances) with a variant of the cavity method
for optimization, specifically the reinforced Max-Sum algorithm. The
MStT has applications in many areas ranging from biology (e.g., finding
protein associations in cell signaling \cite{huang_integrating_2009,bailly-bechet_prize-collecting_2009,tuncbag_simultaneous_2012})
to network technologies (e.g., finding optimal ways to deploy fiber
optic and heating networks for households and industries \cite{ljubic_algorithmic_2005}). 

We will describe the main problem and variants in the rest of this
section. On \prettyref{sec:The-model} we will describe the model
we will use to represent the optimization problems. The formulation
presented here is based on edge variables as opposed to vertex variables
in the ``pointer-depth'' formulation used in \cite{bailly-bechet_prize-collecting_2009,bayati_rigorous_2008,bayati_statistical_2008,biazzo_performance_2012}.
 On both representations, the diameter of the solution tree is bounded
by a parameter that both affects the computation time of each iteration
and is intimately related to convergence properties. However, specially
on 2-dimensional and 3-dimensional instances, the diameter of the
solutions grows much faster with the system size with respect to random
graphs (as a power of the system size instead of the logarithm of
it). We propose an alternative formulation (the \textit{flat} model)
in Subsection \prettyref{sub:The-Flat-Model} that overcomes the \emph{a
priori} bound on the diameter, and that is made possible by the new
edge-variables formulation.

In real-world instances, the de-correlation hypotheses behind the
cavity equations are seldom accurate. The reinforced Max-Sum equations
are partially able to overcome this inaccuracy by using intermediate
results to slowly bootstrap the system into one with large enough
external fields on which the equations are more accurate. However,
in such a slow procedure no solution is found on intermediate steps
before the final convergence, if ever, arrives. This is a big drawback
for real-world optimization, as alternative algorithms, either based
on local search heuristics or linear programming, can on the contrary
provide reasonably good solutions on intermediate steps. In this work
we study the practical problem of obtaining solutions within constricted
time limits, by implementing fast heuristic methods to construct good
trees from partial results while the main computation is in course.
This will be described in \prettyref{sec:A-Belief-Propagation-Inspired}. 

We will present results on low-dimensional and scale-free graphs on
\prettyref{sec:Results-for-Scale-Grid}, and finally describe the
results of the 2014 DIMACS Challenge on \prettyref{sec:Results-for-DIMACS},
in which the algorithm developed in a preliminary version of this
work participated. In order to participate on the challenge, a strategy
to determine the various parameters of the algorithm without human
intervention was devised in order to make the optimization algorithm
fully automatic as per the challenge rules. We consider the results
obtained on the challenge to be very promising, specially considering
the fact that competing algorithms were based on industrial-grade
commercial solver libraries while our approach can be implemented
in a few hundred lines of plain C code.

\subsection{Definition\label{sub:Definition}}

In this section we will define the Rooted Steiner Tree Problem (RSTP)
on graphs.

\paragraph{\emph{The }Rooted Prize-collecting Steiner tree problem. }

Given an undirected graph $G=(V,E)$ with positive real weights $w_{ij}>0$
for $(i,j)\in E$ associated with edges ($w_{ij}$ could be different
from $w_{ji}$), non-negative real prizes $c_{i}$ associated with
$i\in V$ and a single root vertex $r\in V$, we consider the problem
of finding a directed tree $T=\left(V_{T}\subset V,E_{T}\subset E\right)$
rooted at $r$ that minimizes the cost or energy function:
\begin{equation}
\mathcal{H}\left(T\right)=\sum_{\left(i,j\right)\in E_{T}}w_{ij}+\sum_{i\in V\backslash V_{T}}c_{i}\label{eq:ham}
\end{equation}

That is, we seek
\begin{equation}
T^{\star}=\arg\min_{T}\mathcal{H}\left(T\right)\label{eq:minH}
\end{equation}

Without loss of generality, we can assume the minimum in \eqref{eq:minH}
to be unique, by eventually adding negligibly small random noise to
weights $w_{ij}$. Nodes with $c_{i}>0$ are called \emph{profitable
vertices, }or sometimes \emph{generalized terminals} (or even \emph{terminals})
by extension of the classical Steiner Problem on graphs where subset
of \textit{terminal} nodes must be included in the tree.

The following problems can be trivially (polynomially) mapped into
RSTP: Prize-collecting Steiner Tree Problem (PCSPG), Steiner Tree
Problem on graphs (SPG), Group Steiner Problem (GSPG), Minimum Weighted
Steiner Trees (MWM). We will discuss some specific mapping issues
in Section \prettyref{sub:Rooting}.

\section{The model\label{sec:The-model}}

\subsection{Variables}

Consider a rooted (directed) tree $T$, a feasible solution of the RSTP defined in Subsection \prettyref{sub:Definition}, in which the maximum allowed distance between any leaf and the root is parametrized by $D$ (we say that the tree is $D-$bounded). From
$T$, we will define for each oriented $\left(i,j\right)\in E$ an
integer variable $d_{ij}\in\{-D,\ldots,D\}$ as follows. For each
directed edge $\left(i,j\right)$ in $E_{T}$ (directed edges in $E_{T}$
will be assumed to point \emph{to} $r$), $d_{ij}$ will be the distance
(in hops), or \textit{depth}, from $i$ to $r$ along the tree. That
is, $d_{ij}$ is the length of the unique simple path $\left(v_{0}=i,v_{1}=j,\dots,v_{d_{ij}}=r\right)$
in $T$ from $i$ to $r$. For each edge $\left(i,j\right)$ such
that $\left(j,i\right)\in E_{T}$, define $d_{ij}=-d_{ji}$. The sign
of $d_{ij}$ will define unambiguously the orientation of edge $\left(i,j\right)$
in the tree with respect to the root $r$. For edges such that both
$\left(i,j\right),\left(j,i\right)\notin E_{T}$, we define $d_{ij}=0$.
The vector $\mathbf{d}=\left\{ d_{ij}:\left(i,j\right)\in E\right\} $
so defined clearly satisfies an antisymmetric condition $d_{ij}=-d_{ji}$
for each $(i,j)\in E$. Such $\mathbf{d}$ will be henceforth called
a \emph{representation }of $T$. It is easy to verify that the mapping
$T\mapsto\mathbf{d}$ is one-to-one: the inverse mapping is $\mathbf{d}\mapsto\left(V_{\mathbf{d}},E_{\mathbf{d}}\right)$
with $V_{\mathbf{d}}=\left\{ i\in V:\exists k\in V:d_{ki}\neq0\right\} $
and $E_{\mathbf{d}}=\left\{ \left(i,j\right):d_{ij}\neq0\right\} $.

\subsection{Constraints}

We would like to switch to an optimization problem on the vector $\mathbf{d}$
that mirrors the RSTP one. In order to ensure that the $\left(V_{\mathbf{d}},E_{\mathbf{d}}\right)$
are trees, we will need to impose rigid constraints on the vector
$\mathbf{d}$ besides the antisymmetric condition. This will be implemented
as a family of \emph{local} constraints on sub-vectors of variables
$\mathbf{d}_{i}=\left\{ d_{ji}:j\in V\left(i\right)\right\} $ where
the symbol $V\left(i\right)$ will denote the \emph{neighborhood}
of $i$, $V\left(i\right)=\left\{ j\in V:\left(i,j\right)\in E\right\} $.
For each node $i$ we define a proper compatibility function $\psi_{i}(\boldsymbol{d}_{i})$,
that is equal to one if the constraints are satisfied and zero otherwise.
As the sign of $d_{ij}$ represents the orientation of edges along
the tree (with $d_{ij}>0$ if $j$ is closer to $r$ than $i$), two
mutually excluding situations can occur in the neighbor of $i$. Either
$i$ does not belong to $T$, and so $d_{ij}=0$ for each $j\in V\left(i\right)$,
or else there exists exactly one neighbor $k$ such that $d_{ik}>0$,
and for the remaining neighbors $l\in V\left(i\right)\setminus k$,
either $d_{li}=0$ or $d_{li}=d_{ik}+1$. The root node $r$ is special,
as there is no neighbor closer to $r$ than itself; i.e. for each
$k\in V\left(r\right)$, $d_{kr}$ is either $0$ or $1$. Symbolically,
admissible configurations of $\mathbf{d}_{i}$ can be encoded by the
nonzero arguments of the following \emph{compatibility} function:
\begin{eqnarray}
\psi_{i}(\boldsymbol{d}_{i}) & = & \prod_{k\in V\left(i\right)}\delta_{d_{ki},0}+\sum_{d>0}\sum_{k\in V\left(i\right)}\left[\delta_{d_{ki},-d}\prod_{l\in V\left(i\right)\setminus k}(\delta_{d_{li},d+1}+\delta_{d_{li},0})\right]\mbox{ for }i\neq r\label{eq:psi}\\
\psi_{r}(\boldsymbol{d}_{r}) & = & \prod_{k\in V\left(r\right)}(\delta_{d_{kr},1}+\delta_{d_{kr},0})\label{eq:psi2}
\end{eqnarray}

where $\delta$ denotes the Kronecker delta. Once $\psi_{i}$ are
defined, a cost function $\mathcal{H}$ (we will use the same symbols
as the one for trees) can be defined for variables $\mathbf{d}$ such
that vectors $\mathbf{d}$ with finite $\mathcal{H}$$\left(\mathbf{d}\right)$
represent some tree $T$, and in such case, $\mathcal{H}\left(T\right)=\mathcal{H}\left(\mathbf{d}\right)$:
\[
\mathcal{H}\left(\mathbf{d}\right)=\begin{cases}
\infty & \mbox{if }\prod_{i}\psi_{i}\left(\mathbf{d}_{i}\right)=0\\
\sum_{i}\left\{ c_{i}\mathbb{I}\left[\mathbf{d}_{i}\equiv\mathbf{0}\right]+\sum_{j\in V\left(i\right)}w_{ij}\mathbb{I}\left[d_{ij}>0\right]\right\}  & \mbox{if }\prod_{i}\psi_{i}\left(\mathbf{d}_{i}\right)=1
\end{cases}
\]

where $\mathbb{I}$ is the indicator function which takes value $1$
if the argument is true or $0$ otherwise. Then we can substitute
\eqref{eq:minH} with
\[
\mathbf{d}^{\star}=\arg\min_{\mathbf{d}}\mathcal{H}\left(\mathbf{d}\right)
\]

\section{\label{sec:Belief-Propagation}Belief Propagation}

\subsection{Belief Propagation equations}

The Belief Propagation (BP) equations (or Replica-Symmetric Cavity
Equations in Statistical Physics), are a set of equations to describe
approximately a Boltzmann-Gibbs distribution in terms of some of its
marginals. The Boltzmann distribution is in this case a probability
measure on the space of all Steiner Trees on $G$ (represented by
vectors $\mathbf{d}$), where lower cost trees have larger probability
in a way that is dependent on a positive value $\beta$ called the
\emph{inverse temperature}. The corresponding Boltzmann distribution
is in this case: 
\begin{eqnarray}
P\left(\mathbf{d}\right) & = & \frac{1}{Z}e^{-\beta\mathcal{H}\left(\mathbf{d}\right)}\label{eq:Boltzmann1}\\
 & = & \frac{1}{Z}\prod_{i}\psi_{i}\left(\mathbf{d}_{i}\right)e^{-\beta\sum_{i}\left(c_{i}\mathbb{I}\left[\mathbf{d}_{i}\equiv\mathbf{0}\right]+\sum_{j\in V\left(i\right)}w_{ij}\mathbb{I}\left[d_{ij}>0\right]\right)}\label{eq:Boltzmann2}
\end{eqnarray}

where the $\beta$-dependent constant $Z$ (called the partition function)
can be obtained by the normalization condition of the probability
measure $P$. When $\beta\to\infty$, the distribution concentrates
on the optimal tree(s). The marginal function $P\left(d_{ij}\right)$
(defined with a slight abuse of notation using the same symbol $P$)
consists in the quantity
\begin{equation}
P\left(d_{ij}\right)=\sum_{\mathbf{d}'}P\left(\mathbf{d}'\right)\delta_{d_{ij},d_{ij}^{'}}
\end{equation}

Calculating marginals is generally hard in computational terms but
extremely useful; for example, in the $\beta\to\infty$ limit, knowing
the value of a marginal gives crucial information on the state of
variable $d_{ij}$ in the optimal configuration(s). It is easy to
see that this calculation can be used recursively to compute the optimum
itself.

The BP equations are a set of fixed-point equations for \emph{modified
or cavity }marginal functions. They can be shown to become asymptotically
exact for some models on certain types of random graphs in the limit
$\left|V\right|\to\infty.$ On single instances, the Belief Propagation
equations are widely employed for inference in various contexts, including
telecommunication and visual stereo recognition. The equations are
exact on any acyclic graph and correspond to a dynamic programming
solution for the corresponding inference problem. We will not enter
here into details on the nature of the approximation behind the equations;
we refer the interested reader to \cite{mezard_information_2009}. 

The cavity messages for our model are $m_{ij}:\left\{ -D,\dots,D\right\} \mapsto[0,1]$,
also called \textit{messages}, and its general expression \cite{mezard_information_2009}
can be written for this particular model as follows:
\begin{eqnarray}
m_{ij}\left(d_{ij}\right) & \propto & \sum_{\mathbf{d}_{i\setminus j}}\psi_{i}\left(\mathbf{d}_{i}\right)e^{-\beta\left(c_{i}\mathbb{I}\left[\mathbf{d}_{i}\equiv\mathbf{0}\right]+\sum_{k\in V\left(i\right)}w_{ik}\mathbb{I}\left[d_{ik}>0\right]\right)}\prod_{k\in V\left(i\right)\setminus j}m_{ki}\left(d_{ki}\right)\label{eq:BP1}
\end{eqnarray}

where $\mathbf{d}_{i\setminus j}=\left\{ d_{ki}:k\in V\left(i\right)\setminus j\right\} $.
The proportional sign $\propto$ above hides a normalization constant
that can be computed \textit{a posteriori} after the computation of
the rest of the right-hand side of \eqref{eq:BP1} for all values
of $d_{ij}$. Notice that the computation of this constant is fundamentally
easier than the computation of the partition function $Z$ in \eqref{eq:Boltzmann1},
as the corresponding sum involves only $2D+1$ terms rather than an
exponential number. On a fixed point, an approximation for the marginals
$P\left(d_{ij}\right)$ is given by
\begin{equation}
P\left(d_{ij}\right)\propto m_{ij}\left(d_{ij}\right)m_{ji}\left(-d_{ij}\right)\label{eq:BP2}
\end{equation}

The equations are normally employed as follows: a numerical solution
for the fixed-point equations \eqref{eq:BP1} is sought by iteration
from a random initial condition, and, on the fixed point, \eqref{eq:BP2}
is applied to obtain (approximated) marginals.

\subsection{The $\beta\to\infty$ limit: Max-Sum equations}

As we are interested in the optimization problem, we will take the
$\beta\to\infty$ limit of \eqref{eq:Boltzmann1}-\eqref{eq:Boltzmann2}.
As standard (see e.g. \cite{mezard_information_2009}), one performs
a change of variables into \emph{cavity fields} $h_{ij}\left(d_{ij}\right)=\frac{1}{\beta}\log m_{ij}\left(d_{ij}\right)$
and \emph{local fields} $H_{ij}\left(d_{ij}\right)=\frac{1}{\beta}\log P\left(d_{ij}\right)$.
For $\beta\to\infty$, local fields\emph{ }$H_{ij}$ give very valuable
information about locally restricted optima:
\[
H_{ij}\left(d_{ij}\right)=\min_{\substack{\mathbf{d}':\psi_{i}\left(\mathbf{d'}\right)\equiv1}
}\mathcal{H}\left(\mathbf{d}'\right)-\min_{\substack{\mathbf{d}':\psi_{i}\left(\mathbf{d'}\right)\equiv1\\
d'_{ij}=d{}_{ij}
}
}\mathcal{H}\left(\mathbf{d}'\right)
\]
from which a global optimum can be easily computed. We substitute
the change of variables into $\eqref{eq:BP1}-\eqref{eq:BP2}$, to
obtain in the \emph{$\beta\to\infty$} limit the Max-Sum (MS) equations:
\begin{eqnarray}
h_{ij}\left(d_{ij}\right) & = & \max_{\substack{\mathbf{d}_{i\setminus j}\\
\psi_{i}\left(\mathbf{d}_{i}\right)=1
}
}\bigg\{-c_{i}\mathbb{I}\left[\mathbf{d}_{i}\equiv\mathbf{0}\right]+\nonumber \\
 &  & \;-\sum_{k\in V\left(i\right)}w_{ik}\mathbb{I}\left[d_{ik}>0\right]+\sum_{k\in V\left(i\right)\setminus j}h_{ki}\left(d_{ki}\right)\bigg\}+C\label{eq:MS1}\\
H_{ij}\left(d_{ij}\right) & = & h_{ij}\left(d_{ij}\right)+h_{ji}\left(-d_{ij}\right)+C'\label{eq:MS2}
\end{eqnarray}
where $C,\,C'$ are additive constants (i.e., that do not depend on
$d_{ij}$) that can be computed after the rest of the right-hand side,
and ensure that $\max_{d}h_{ij}\left(d\right)=\max_{d}H_{ij}\left(d\right)=0$;
such condition corresponds to the normalization constraint on messages
$m_{ij}$ and marginals $P_{ij}$ for finite $\beta$. For shortness,
we will drop from now on the additive constants in the equations. 

As with BP, MS equations are normally solved numerically by repeated
iteration of \eqref{eq:MS1}. On a fixed point, \eqref{eq:MS2} is
computed and then, for each $H_{ij}$, we perform the maximum over
$d_{ij}$
\begin{equation}
d_{ij}^{\star}=\arg\max H_{ij}\left(d_{ij}\right).\label{eq:cavmax}
\end{equation}

If this maximum in \prettyref{eq:cavmax} is unique, a tree can be
reconstructed by using the inverse mapping defined in \prettyref{sec:The-model}.
Variables $d_{ij}^{\star}$ are called \emph{decisional variables}.

Notably, it can be shown \cite{bayati_rigorous_2008,biazzo_performance_2012}
that MS equations are exact on \emph{arbitrary} graphs for the Spanning
Tree problem, i.e. when a positive large enough constant $c$ is associated
with each node $i\in V$. If a fixed point is found and the maximum
$d_{ij}^{\star}$ are non degenerate (i.e. the maximums are unique),
then they form the representation of the (then unique) Minimum Spanning
Tree. The degeneracy requirement can be relaxed by adding to edge
weights random noise terms $r_{ij}$, negligibly small with respect
to $w_{ij}$.

\subsection{Reinforcement}

For the Minimum Steiner Tree or Prize-collecting Steiner Tree Problem,
iteration of Equations \eqref{eq:MS1}-\eqref{eq:MS2} very seldom
converge. Nevertheless, there is still valuable information in local
fields before convergence. The following strategy can be applied \cite{bayati_statistical_2008,bailly-bechet_finding_2011,biazzo_performance_2012}:
add a reinforcement term to \eqref{eq:MS1}, that progressively \emph{bootstraps}
the model into an easy one in the direction of a feasible configuration.

The dynamical equations are:
\begin{eqnarray}
h_{ij}^{t+1}\left(d_{ij}\right) & = & \max_{\substack{\mathbf{d}_{i\setminus j}\\
\psi_{i}\left(\mathbf{d}_{i}\right)=1
}
}\left\{ -c_{i}\mathbb{I}\left[\mathbf{d}_{i}\equiv\mathbf{0}\right]-\sum_{k\in V\left(i\right)}w_{ik}\mathbb{I}\left[d_{ik}>0\right]+\sum_{k\in V\left(i\right)\setminus j}h_{ki}^{t}\left(d_{ki}\right)\right\} \label{eq:MS3}\\
H_{ij}^{t+1}\left(d_{ij}\right) & = & h_{ij}^{t+1}\left(d_{ij}\right)+h_{ji}^{t+1}\left(-d_{ij}\right)+\gamma_{t}H_{ij}^{t}\left(d_{ij}\right)\label{eq:MS4}
\end{eqnarray}

where $\gamma_{t}$ is called the \emph{reinforcement factor}. The
factor $\gamma_{t}$ is increased in a linear regime $\gamma_{t}=\gamma_{1}t$,
where the parameter $\gamma_{1}$ is typically small, for instance taken in the interval $[10^{-5},10^{-3}]$.
Rather than waiting for convergence of the equations, decisional variables
$d_{ij}^{\star}$ are computed during iterations, and the process
is stopped when decisional variables are repeated a predefined number
of times (e.g. 50--100). The number of iteration needed is proportional
to $\gamma_{1}^{-1}$, as it is observed that the process generally
stops when $\gamma_{t}\simeq1$. For best results, $\gamma_{1}$ should
be small (so the bootstrapping procedure is slow) and it is crucial
to have an extremely efficient computation of the $2D\left|E\right|$
values on the left of \prettyref{eq:MS3}. As we will see, these can
be computed in time $\Theta\left(D\left|E\right|\right)$.

\subsection{Efficient computation of the equations\label{sub:Efficient-computation}}

Equation \eqref{eq:MS3} requires the computation of a maximum over
a set of $\left(2D+1\right)^{\left|V\left(i\right)\right|}$ elements,
that quickly becomes prohibitive even for modest values of $D$ and
$\left|V\left(i\right)\right|$. Fortunately, the computation can
be performed in amortized $O\left(2D+1\right)$ time as follows. First,
let us consider the root node $r$. In this case the compatibility
function is particularly simple: neighboring edges can be present
with $d_{kr}=1$ or absent with $d_{kr}=0$ in the tree. The equations
consequently simplify enormously: 
\begin{eqnarray}
h_{rj}^{t+1}\left(d_{rj}\right) & = & \begin{cases}
\sum_{k\in V\left(r\right)\setminus j}\max\left\{ h_{kr}^{t}\left(1\right),\,h_{kr}^{t}\left(0\right)\right\}  & \mbox{ for }d_{rj}=-1,0\\
-\infty & \mbox{ for }d_{rj}\neq-1,0
\end{cases}
\end{eqnarray}

Let us consider any $i\neq r$. Suppose first $d_{ij}>0$. Then, in
order for $\psi_{i}$ to be $1$, it must be that $d_{ki}=d_{ij}+1$
or $d_{ki}=0$ for each $k\in V\left(i\right)\setminus j$: 
\begin{eqnarray}
h_{ij}^{t+1}\left(d_{ij}\right) & = & -w_{ij}+\sum_{k\in V\left(i\right)\setminus j}\max\left\{ h_{ki}^{t}\left(d_{ij}+1\right),\,h_{ki}^{t}\left(0\right)\right\} \\
 & = & -w_{ij}+\sum_{k\in V\left(i\right)}\max\left\{ h_{ki}^{t}\left(d_{ij}+1\right),\,h_{ki}^{t}\left(0\right)\right\} +\\
 &  & -\max\left\{ h_{ji}^{t}\left(d_{ij}+1\right),\,h_{ji}^{t}\left(0\right)\right\} 
\end{eqnarray}

The above equation can be clearly computed for all $j\in V\left(i\right)$
in $\Theta\left(D\left|V\left(i\right)\right|\right)$ operations
(first computing the sum, then subtracting one term for each neighbor).
For $i\neq r$, $d_{ij}=-1$ is forbidden, so $h_{ij}^{t+1}\left(-1\right)=-\infty$.
Suppose then $d_{ij}<-1$. In this case there must exist $k\in V\left(i\right)\setminus j$
such that $d_{ik}+1=d_{ji}$, and all other $l\in V\left(i\right)\setminus\left\{ k,\,j\right\} $,
$d_{li}=-d_{ij}$ or $d_{li}=0$. In symbols, 
\begin{eqnarray}
h_{ij}^{t+1}\left(d_{ij}\right) & = & \max_{k\in V\left(i\right)\setminus j}\big\{ h_{ki}^{t}\left(d_{ij}+1\right)-w_{ik}+\sum_{l\in V\left(i\right)\setminus k,j}\max\left\{ h_{li}^{t}\left(0\right),\,h_{li}^{t}\left(-d_{ij}\right)\right\} \big\}\\
 & = & \sum_{l\in V\left(i\right)}\max\left\{ h_{li}^{t}\left(0\right),\,h_{li}^{t}\left(-d_{ij}\right)\right\} -\max\left\{ h_{ji}^{t}\left(0\right),\,h_{ji}^{t}\left(-d_{ij}\right)\right\} +\nonumber \\
 &  & +\max_{k\in V\left(i\right)\setminus j}\left\{ h_{ki}^{t}\left(d_{ij}+1\right)-w_{ik}-\max\left\{ h_{ki}^{t}\left(0\right),\,h_{ki}^{t}\left(-d_{ij}\right)\right\} \right\} \label{eq:cavitymax}
\end{eqnarray}

Note that $A=\max_{k\in V\left(i\right)\setminus j}A_{k}$, where
$A_{k}=h_{ki}^{t}\left(d_{ij}+1\right)-w_{ik}-\max\left\{ h_{ki}^{t}\left(0\right),\,h_{ki}^{t}\left(-d_{ij}\right)\right\} $
in \eqref{eq:cavitymax} can be computed in $\left|V\left(i\right)\right|$
operations for all $j\in V\left(i\right)$. First, in $\Theta\left(\left|V\left(i\right)\right|\right)$
operations, the first two maxima $A_{k_{1}},\,A_{k_{2}}$ can be computed
along with $k_{1},$ the position of the first maximum. Then, for
$j\in V\left(i\right)$, if $j=k_{1},$ then $A=A_{k_{2}}$and for
$j\neq k_{1}$, $A=A_{k_{1}}$.

Finally, the case $d_{ij}=0$ is similar to the one with $d_{ij}<-1$
and can be computed by reusing the computation above:
\begin{eqnarray}
h_{ij}^{t+1}\left(0\right) & = & \max\left\{ \sum_{k\in V\left(i\right)\setminus j}h_{ki}^{t}\left(0\right)-c_{i},\:\max_{d<-1}h_{ij}^{t+1}\left(d\right)\right\} 
\end{eqnarray}

In summary, \eqref{eq:MS3} can be computed for all neighbors of a
vertex $i$ in a time proportional to $D\left|V\left(i\right)\right|$,
which gives a total time per iteration proportional to $D\left|E\right|$.
As this is obviously the same time requirement of \eqref{eq:MS4},
each iteration of the MS equations requires a number of elementary
operations proportional to $D\left|E\right|$.

\subsection{The \emph{Flat Model\label{sub:The-Flat-Model}}}

For variants of the problem in which the hop-length from the root
is unlimited, variables $d_{ij}$ should be unbounded. Fortunately,
if $d_{ij}\in\left\{ -D,\dots,D\right\} $, a value of $D=\left|V\right|$
would be clearly sufficient. However, the computation of the equations
scales in time as $D\left|E\right|$; so it is generally not desirable
to allow too large values of $D$. As we will see, a slightly modified
model allows to significantly reduce the needed depth bound $D$ to
$D=\left|K\right|$ where $K$ is the set of \emph{generalized terminals,
}i.e. nodes with $c_{i}>0$. The idea is to allow in the representation
$\mathbf{d}$ of the tree, to have chains of edges with identical
depth $d$, i.e. $v_{0},\dots,v_{k}$ with $d_{v_{0}v_{1}}=d_{v_{1}v_{2}}=\cdots=d_{v_{k-1}v_{k}}$.
We will allow this situation (the \emph{flat} \emph{rule}, to differentiate
it from the \emph{normal} rule \eqref{eq:psi}) for a node $v_{s}$
only if: (i) $v_{s}$ is not a terminal and (ii) $v_{s}$ has degree
exactly two in the tree (i.e. no other neighbor besides $v_{s-1}$
and $v_{s+1}$). These two conditions ensure that (optimal) configurations
satisfying this relaxed set of constraints represent trees; extra
cycles with identical depth, containing no terminal, can of course
be present, but are suboptimal in terms of cost. In symbols, we would
use instead of the compatibility function in \eqref{eq:psi},
\begin{eqnarray}
\psi'_{i}\left(\mathbf{d}_{i}\right) & = & \psi_{i}\left(\mathbf{d}_{i}\right)+\psi_{i}^{flat}\left(\mathbf{d}_{i}\right)\label{eq:psi_flat}\\
\psi_{i}^{flat}\left(\mathbf{d}_{i}\right) & = & \delta_{c_{i},0}\sum_{d>0}\sum_{k\in V\left(i\right)}\sum_{l\in V\left(i\right)\setminus k}\delta_{d_{ki},-d}\delta_{d_{li},d}\prod_{j\in V\left(i\right)\backslash k,l}\delta_{d_{ij},0}
\end{eqnarray}

Two remarks are in order: the correspondence between a tree $T$ and
a representation $\mathbf{d}$ is no more one-to-one: different vectors
$\mathbf{d}$ represent the same tree $T$ (because in a non-branching
path inside $T$, depth can either increase or not increase). Moreover,
as we have seen above, some allowed configurations $\mathbf{d}$ now
do not represent any tree (because they may have extraneous disconnected
cycles). Nevertheless, such apparent inconsistencies are not problematic.
To see this, consider the following two statements:
\begin{enumerate}
\item Given $\mathbf{d}$ such that $\psi'_{i}(\boldsymbol{d}_{i})\equiv1$,
consider $S=\left(V_{\mathbf{d}},E_{\mathbf{d}}\right)$ with $E_{\mathbf{d}}=\left\{ \left(i,j\right):d_{ij}\neq0\right\} $,
$V_{\mathbf{d}}=\left\{ i\in V:\exists k\in V:d_{ki}\neq0\right\} $.
Then the graph $S$ consists of the disjoint union of a tree $T$
and zero or more disconnected components that are simple cycles that
do not own any terminal.

Along any directed cycle, depth cannot decrease (otherwise along the
path we would find a vertex $j$ with $d_{ij}<d_{jk}$ which is forbidden
both by the normal and the \emph{flat} rule); therefore only the flat
rule could have been employed. Thus, there cannot exist any branching
vertex (nor terminal, including the root node $r$) in the cycle and
the depth should be constant. This cycle with constant depth will
form a separate connected component with no profitable vertices. The
connected component of $r$, on the other hand, cannot have cycles,
and being it acyclic and connected, it is a tree.\label{enu:prop1}

\item Any optimal tree has a \emph{flat} representation $\mathbf{d}$ with
$D=\left|K\right|$ where $K=\left\{ v\neq r:c_{v}>0\right\} $.

For unbounded $D$, consider $\mathbf{d}$ the most compact representation
for an optimal tree $\mathcal{T}$, i.e. no depth increases in any
non-branching, non-terminal vertex on the tree. Consider the unique
simple path $(r=v_{0},\dots,v_{k}=l)$ from the root to a leaf in
$\mathcal{T}$. We will prove that $d_{v_{i+1},v_{i}}\leq1+K-K_{i}$
for $i=0,\dots k-1$, where $K_{i}$ is the number of terminals different
from $i$ in the subtree $\mathcal{T}_{i}$ rooted at $i$. As in
an optimal tree, every leaf should be a terminal, the proposition
would be proved by considering $i=k-1$, as $d_{v_{k}v_{k-1}}\leq1+K-1=K$.
Let us prove it by induction on $i$. The result is clearly true for
$i=0$, as $d_{v_{1}r}=1\leq1+K-K_{i}=1$. Suppose the result true
for some $0\leq i<k-1$. There are two cases to consider: 1) $d_{v_{i+2}v_{i+1}}=d_{v_{i+1}v_{i}}$.
Then $v_{i+1}$ is non-branching and $c_{v_{i+1}}=0$ (otherwise the
\textit{flat} rule cannot be applied). But then $K_{i+1}=K_{i}$ (because
the tree $\mathcal{T}_{i}$ is just $\mathcal{T}_{i+1}$ plus the
edge $\left(i,i+1\right)$ and $i+1$ is non-terminal). Clearly then
$d_{v_{i+2}v_{i+1}}=d_{v_{i+1}v_{i}}\leq1+K-K_{i}=1+K-K_{i+1}$. 2)
Either $v_{i}$ is terminal or $v_{i}$ is \emph{branching }in $\mathcal{T}$\emph{,
}i.e. $v_{i}$ has at least one children $j$ different from $v_{i+1}$
in $\mathcal{T}$. In both cases $K_{i+1}\leq K_{i}-1$: in first
case, simply because $v_{i}$ is a terminal. In the second one because
the subtree $\mathcal{T}_{j}$ will necessarily have at least one
leaf that must be a terminal. Then $d_{v_{i+2}v_{i+1}}=1+d_{v_{i+1}v_{i}}\leq2+K-K_{i}\leq1+K-K_{i+1}$.\label{enu:prop2}

\end{enumerate}
Thus by \eqref{enu:prop1}-\eqref{enu:prop2} we can deduce that the
\emph{flat} mode is not less convenient than the normal model in terms
of solutions, i.e. 
\[
\min_{\mathbf{d}:\psi'_{i}\left(\mathbf{d}\right)\equiv1}\mathcal{H}\left(\mathbf{d}\right)\leq\min_{\mathbf{d}:\psi{}_{i}\left(\mathbf{d}\right)\equiv1}\mathcal{H}\left(\mathbf{d}\right)
\]

and also that for $D\geq\left|K\right|$, the flat model is complete,
i.e.
\begin{equation}
\min_{\mathbf{d}:\psi'_{i}\left(\mathbf{d}\right)\equiv1}\mathcal{H}\left(\mathbf{d}\right)=\min_{T\mbox{ tree}}\mathcal{H}\left(T\right)
\end{equation}
and $\mathbf{d}^{\star}=\arg\min\mathcal{H}\left(\mathbf{d}\right)$
\emph{is a representation of the optimal tree}. 

\begin{figure}[H]
\begin{raggedright}
\includegraphics[width=0.5\columnwidth]{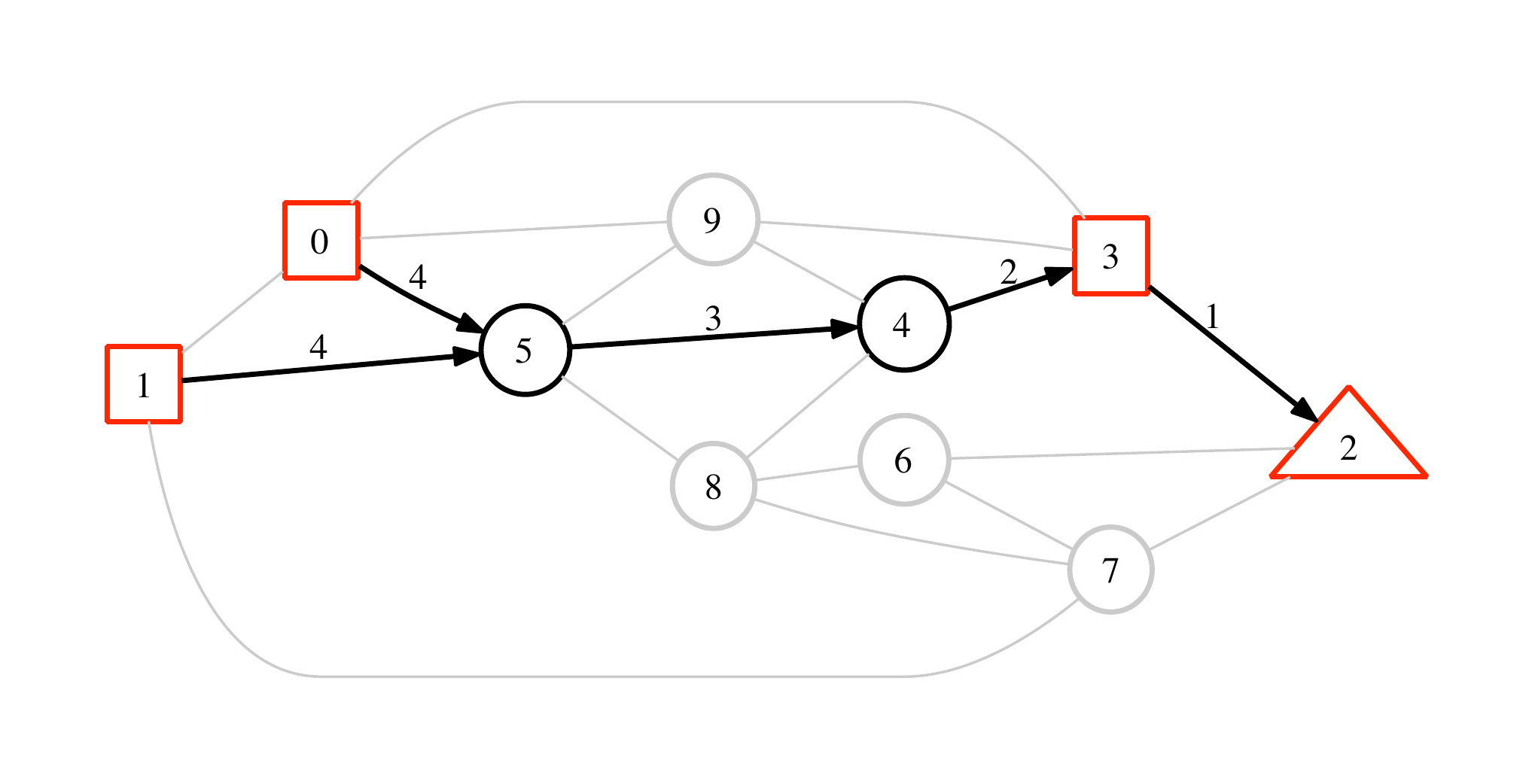}\includegraphics[width=0.5\columnwidth]{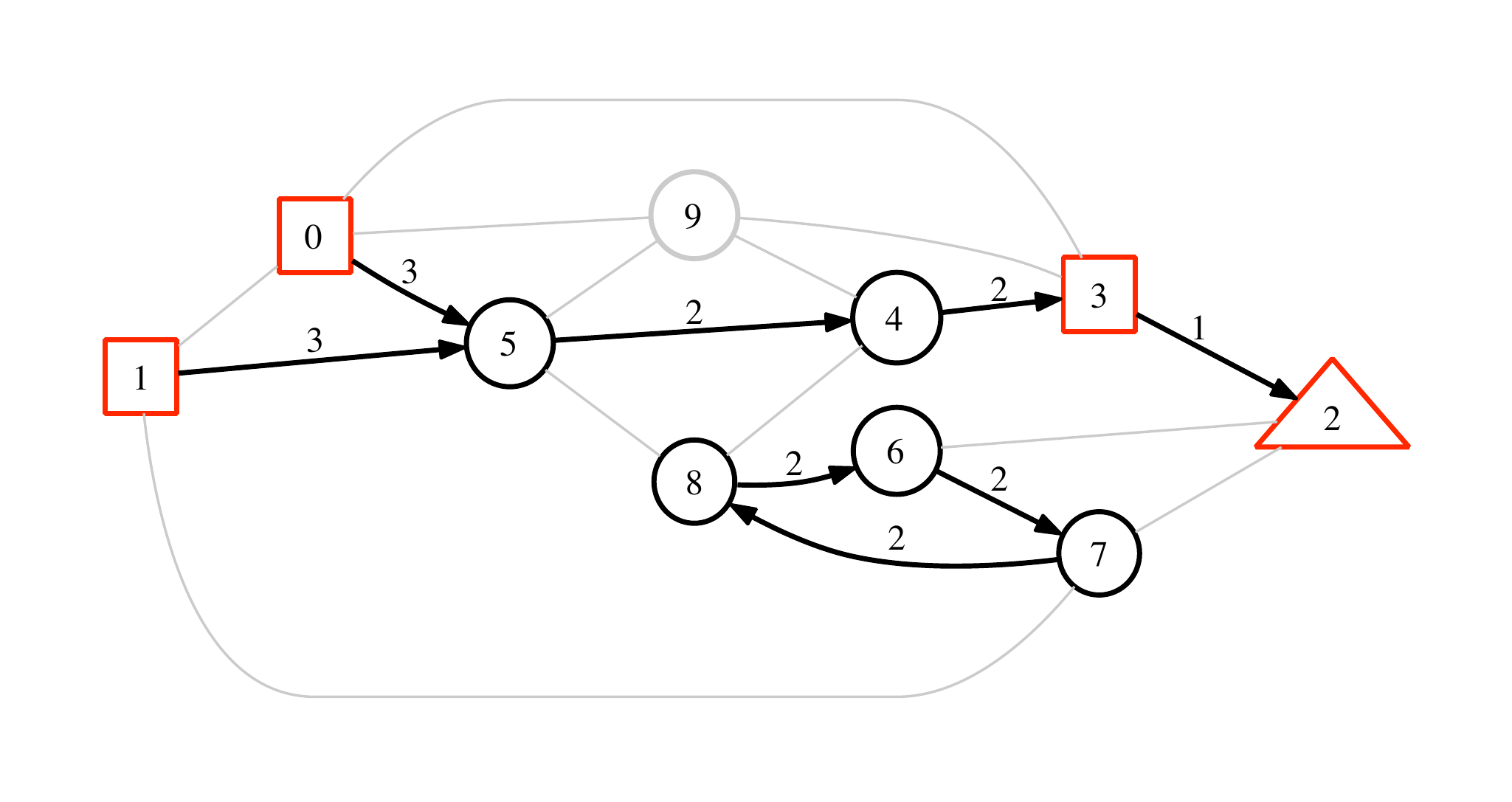}
\par\end{raggedright}

\caption{Left: A feasible solution of the RSTP for a small graph of ten nodes
in the normal representation. Right: A feasible configuration for
the flat representation. \label{fig:Optree}}
\end{figure}

To give an example we plot in Fig. \eqref{fig:Optree} a feasible
solution of the RSTP for a small graph in the $branching$ and in
the $flat$ representation. For sake of simplicity terminals are displayed
as squares, the root as a triangle, and numbers close to edges represent
the decisional variables different from zero. In the normal model,
starting from leaves \{0,1\}, all hop-lengths decrease if we make
a step in the direction of the root, in this case from left to right.
In the flat representation, proceeding from the root ``2'', we first
encounter node ``3'' and, since it is a terminal, so the depth must
increase. After that, being ``4'' neither terminal nor a ``branching
node'' and having degree exactly two within the tree, the depth can
remain equal to 2. Node ``5'' is not a terminal but has degree 3
and so we must increase the depth on children branches. Nodes \{6,
7, 8\} are not terminals and in the cycle of Fig. \eqref{fig:Optree}
have degree 2; thus this disconnected component can be present in
a feasible configuration for the flat model but is energetically inconvenient
and it will be discarded by minimization of the energy.

Finally, the computation for the corresponding \eqref{eq:MS3} for
the \textit{flat} model can be also carried out in overall time proportional
to $D\left|E\right|$ per iteration. The MS equation is
\begin{eqnarray}
h_{ij}^{t+1}\left(d_{ij}\right) & = & \max_{\substack{\mathbf{d}_{i\setminus j}:\\
\psi'_{i}\left(\mathbf{d}_{i}\right)=1
}
}\big\{-c_{i}\mathbb{I}\left[\mathbf{d}_{i}\equiv\mathbf{0}\right]-\sum_{k\in V\left(i\right)}w_{ik}\mathbb{I}\left[d_{ik}>0\right]+\sum_{k\in V\left(i\right)\setminus j}h_{ki}^{t}\left(d_{ki}\right)\big\}\label{eq:MSflat}\\
 & = & \max\left\{ M_{ij}\left(d_{ij}\right),M_{ij}^{flat}\left(d_{ij}\right)\right\} 
\end{eqnarray}

where
\begin{eqnarray}
M_{ij}\left(d_{ij}\right) & = & \max_{\substack{\mathbf{d}_{i\setminus j}:\\
\psi_{i}\left(\mathbf{d}_{i}\right)=1
}
}\left\{ -c_{i}\mathbb{I}\left[\mathbf{d}_{i}\equiv\mathbf{0}\right]-\sum_{k\in V\left(i\right)}w_{ik}\mathbb{I}\left[d_{ik}>0\right]+\sum_{k\in V\left(i\right)\setminus j}h_{ki}^{t}\left(d_{ki}\right)\right\} \\
M_{ij}^{flat}\left(d_{ij}\right) & = & -w_{ij}\mathbb{I}\left[d_{ij}>0\right]+\max_{\substack{\mathbf{d}_{i\setminus j}:\\
\psi_{i}^{flat}\left(\mathbf{d}_{i}\right)=1
}
}\sum_{k\in V\left(i\right)\setminus j}\left\{ h_{ki}^{t}\left(d_{ki}\right)-w_{ik}\mathbb{I}\left[d_{ik}>0\right]\right\} 
\end{eqnarray}

The term $M$ corresponds to the MS equation for the normal model
that can be computed as described in Subsection \ref{sub:Efficient-computation}.
For $i$ such that $c_{i}=0$, the term $M^{flat}$ will be computed
as follows. For $d_{ij}\neq0$
\begin{equation}
\begin{cases}
M_{ij}^{flat}\left(d_{ij}\right)=-w_{ij}+\max_{k\in V\left(i\right)\setminus j}\left\{ h_{ki}^{t}\left(d_{ij}\right)+\sum_{l\in V\left(i\right)\backslash j,k}h_{li}^{t}(0)\right\}  & \quad\mbox{ for }d_{ij}>0\\
M_{ij}^{flat}\left(d_{ij}\right)=\max_{k\in V\left(i\right)\setminus j}\left\{ h_{ki}^{t}\left(d_{ij}\right)-w_{ik}+\sum_{l\in V\left(i\right)\backslash j,k}h_{li}^{t}(0)\right\}  & \quad\mbox{ for }d_{ij}<0
\end{cases}\label{eq:Mflat}
\end{equation}

The restricted maxima in \prettyref{eq:Mflat} for all neighbors $j\in V\left(i\right)$
can be computed in time proportional to $\left|V\left(i\right)\right|$
as in \prettyref{eq:cavitymax}. For $d_{ij}=0$ instead,
\begin{equation}
M_{ij}^{flat}\left(0\right)=\max_{d>0}\max_{\substack{k,\,l\in V\left(i\right)\setminus j\\
k\neq l
}
}\bigg\{ h_{ki}^{t}\left(d\right)+h_{li}^{t}\left(-d\right)-w_{il}+\sum_{m\in V\left(i\right)\backslash j,k,l}h_{mi}^{t}(0)\bigg\}
\end{equation}

For each $d>0$, the internal $\max$ can be computed for all $j\in V\left(i\right)$
again in time proportional $\left|V\left(i\right)\right|$, in a way
similar to the one described in \prettyref{eq:cavitymax} but this
time recording the first three maximums of the quantities in braces
instead of first two. In summary, also using the \textit{flat} rule
the number of needed elementary operations per iteration is $\Theta\left(D\left|E\right|\right)$.

\section{A Belief Propagation-Inspired Heuristics\label{sec:A-Belief-Propagation-Inspired}}

\subsection{Pruned Trees \label{sub:Pruned-Trees}}

One drawback of the MS heuristics with respect to local search based
ones consists in the fact that until convergence of the algorithm,
decisional variables are in a state of inconsistency, incompatible
with a single feasible solution of the problem. A way of obtaining
feasible trees after a few iterations, but long before convergence,
is to use simple heuristics for the SPG and for the PCSPG that use
information carried by MS fields instead of the original edge and
node weights. This procedure will be carefully designed so that in
case of convergence of the MS equations the outcome of the heuristics
is identical to the solution given by the decisional variables of
the MS solution. These strategies become helpful and decisive when
we deal with limited available time, especially for very large instances,
and of course in cases in which plain MS equations do not converge. 

\begin{figure}[H]
\begin{centering}
\includegraphics[width=0.5\columnwidth]{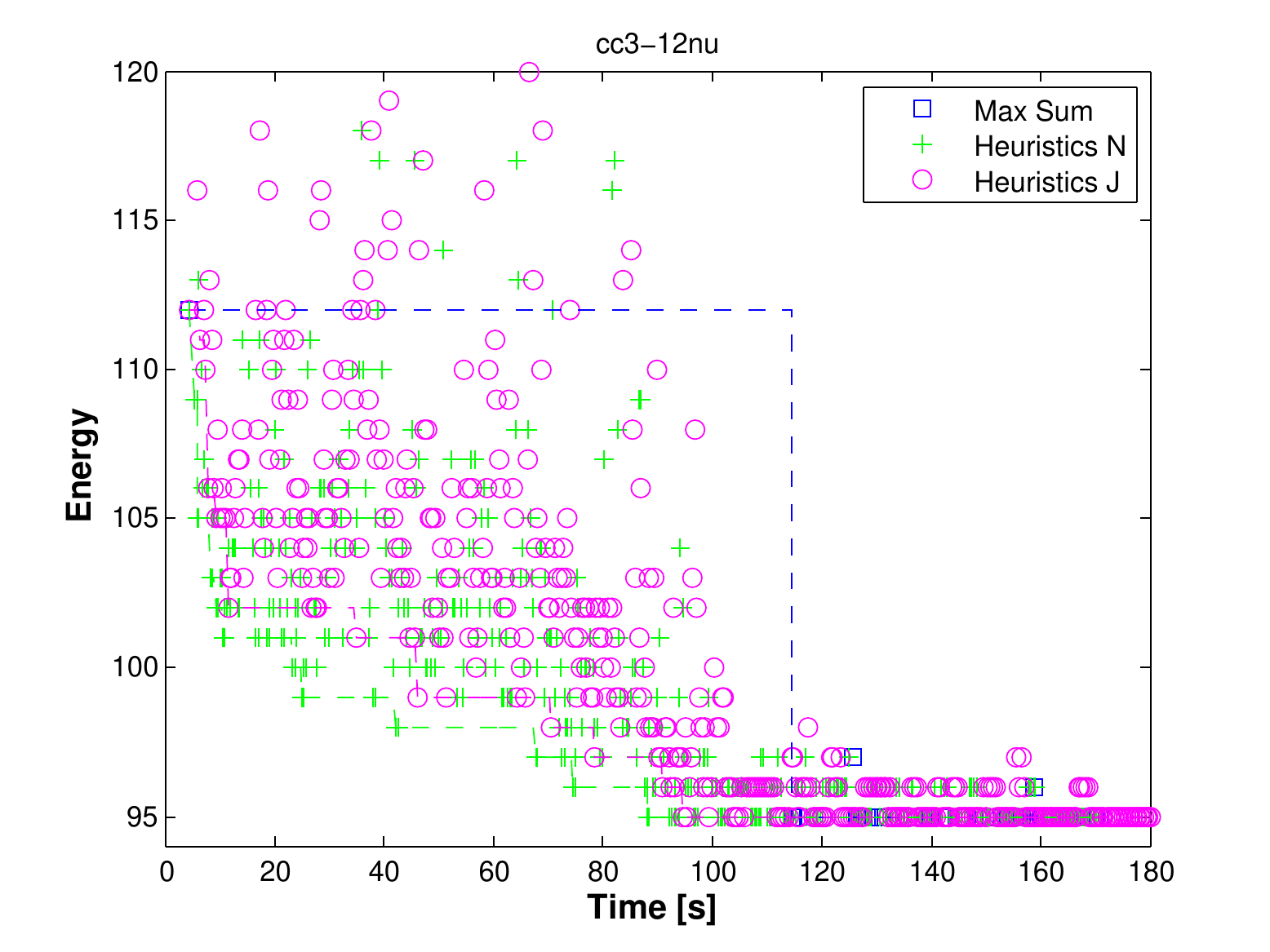}\includegraphics[width=0.5\columnwidth]{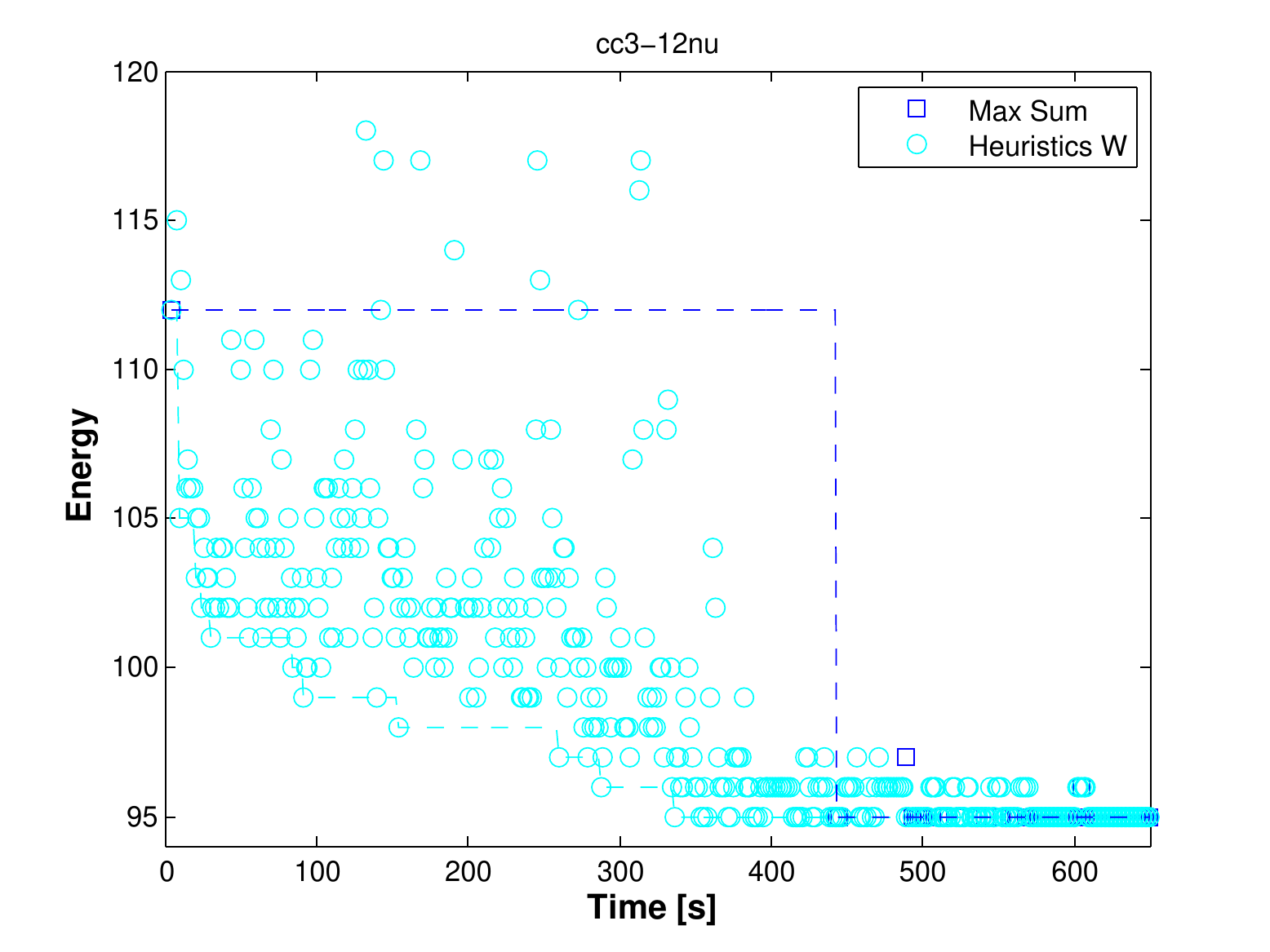}
\par\end{centering}

\caption{Energy of the solution to the PCSPG as function of time on instance
\textit{cc3-12nu\label{fig:Heur_MS}}}
\end{figure}

To give an example we plot in Fig. \eqref{fig:Heur_MS} the outcomes
of the heuristics (labelled as ``N'', ``J'' and ``W'' which
are described in Subsection \ref{sub:Labelling}) and MS, before convergence
of the algorithm, for one of the instances of 11th DIMACS Implementation Challenge, the \textit{cc3-12nu} instance. Points represent
feasible solutions to the PCSPG at any time while we trace the minimum
of the energy provided by each variant using a dashed line. At time
zero we plot the energy of the trivial solution in which the tree
only contains its root. We see that after few hundreds of seconds
the heuristics can give a more energetically favored solution which
improves in time until it coincides with the MS energy at convergence.

\subsubsection{Minimum Spanning Tree and Shortest Path Tree }

At each iteration $t$ we compute a set of modified weights $w_{ij}^{t}$
of the graph $G=(V,\,E)$ which are functions of the local fields
$H_{ij}^{t}$ or cavity fields $h_{ij}^{t}$. We build a feasible
Steiner tree on this re-weighted graph as follows. First, we compute
a spanning tree $T_{H}(V_{H},E_{H})$ (using either \textbf{(a)} Minimum
Spanning Tree (MST) or \textbf{(b)} the Shortest Path Tree (SPT) by
Prim or Dijkstra's algorithm respectively). Afterwards, we apply the
following pruning procedure: starting from each leaf node $i\in V_{H}$
with $V(i)=\{j\}$, we check whether $w_{ij}>c_{i}$. In this case
adding node $i$ to the solution is energetically unfavorable and
we delete $i$ and $(i,j)$ from $T_{H}$. We recursively repeat this
procedure until no such leaf is found. Weights $w_{ij}^{t}$ will
be computed in two alternative ways:
\begin{enumerate}
\item \textit{Reweighting edges}. A first way uses only information contained
in cavity fields $H_{ij}^{t}$: we set $w_{ij}^{t}=\max_{d\neq0}H_{ij}^{t}\left(d\right)$.
This quantity will be strictly positive if the decisional variable
$d_{ij}^{\star}=0$ and will be zero if $d_{ij}^{\star}\neq0$. \label{enu:depth_edges} 
\item \textit{Reweighting nodes}. A second way of assigning auxiliary costs
to edges takes into account the prediction of MS regarding the presence
of each vertex $i$ in the solution. From the equations, a decisional
variable can be assigned to the presence of node $i$ at depth $d\geq0$
by setting
\begin{eqnarray}
h_{i}\left(d\right) & = & \max_{k\in V(i)}\left\{ h_{ik}^{t}(-d)+\sum_{l\in V(i)\setminus k}\max\left\{ h_{li}^{t}\left(d+1\right),h_{li}^{t}\left(0\right)\right\} \right\} \quad\mbox{ for }d>0\nonumber \\
h_{i}\left(0\right) & = & \sum_{k\in V(i)}h_{ki}^{t}(0)-c_{i}\label{eq:h_i}
\end{eqnarray}
We will thus force the presence of nodes $i$ such that $\max_{d>0}h_{i}\left(d\right)>h_{i}\left(0\right)$,
by adding a large prize $C$ to edges connecting nodes not satisfying
this property. \label{enu:depth_nodes}
\end{enumerate}

\subsubsection{Goemans-Williamson heuristics \label{sub:Goemans-Williamson-heuristics}}

For the PCSPG, in addition to the MST and the SPT, we implement the
Goemans-Williamson (GW) algorithm. For the theoretical reasoning and
a detailed description see \cite{Goemans:1992:GAT:139404.139468}
and \cite{Goemans:1996:PMA:241938.241942}. Here we briefly explain
the main steps of the algorithm and how we modify the weights and
the prizes of the graph in order to include the (partial) MS result
within the heuristics.

The algorithm consists in two steps, the ``growth'' stage and the
``pruning'' stage. In the first one we partition vertices in clusters
that are merged and ignored during the iterations until one significant
cluster remains; the more a cluster contains profitable and low cost
connected nodes, the more the cluster will have the chance of being
the final one. The second stage finds a solution of minimum energy
for the PCSPG within the nodes of the final cluster. 

Before applying the algorithm we modify prizes and weights in the
following way. For each node we compute $h_{i}=\max_{d}h_{i}(d)\,-h_{i}(0)$
defined in \eqref{eq:h_i}. If $h_{i}$ is positive, node $i$ is
present in the intermediate solution and so we increase the prize
$c_{i}$ of a large constant $C$; otherwise it keeps its original
prize. In this way we favor those clusters containing nodes with zero
original prize but predicted by MS as Steiner nodes. Moreover, we
modify edges connecting nodes $i:\,h_{i}<0$ by adding $C$ to the
original weight so that clusters whose members are not included in
MS solution are penalized.

\subsection{Labeling \label{sub:Labelling}}

Several approaches and techniques have been proposed in Subsection
\ref{sub:Pruned-Trees}, most of them were not present in the original
algorithm that competed in DIMACS challenge. Experiments are labelled
depending on which model, heuristics and assignment of weights and/or
prizes have been used. All the features of the final algorithm correspond
to the following labels:
\begin{itemize}
\item ``O'': this is the original version of the algorithm which competed
in the challenge and appear in the official results as ``polito''.
It consists in the Max-Sum algorithm for the normal model joined to
the MST; weights are modified as described in \textit{Reweighting
edges} \vpageref{enu:depth_edges}.
\item ``N'': we implement the MST heuristics in which weights are computed
according to \textit{Reweighting nodes} \vpageref{enu:depth_nodes}.
\item ``J'': here we use the SPT heuristics and weights are modified as
in \textit{Reweighting edges} \vpageref{enu:depth_edges}. 
\item ``W'': the heuristics is the GW reported in Subsection \ref{sub:Goemans-Williamson-heuristics}.
\item ``F'': we use the \textit{flat} model described in Subsection \ref{sub:The-Flat-Model}.
If no additional labels are included, we refer to the MST heuristics
with modified weights as in \textit{Reweighting edges} \vpageref{enu:depth_edges}. 
\end{itemize}
Labels can be also combined, e.g. ``F J'' corresponds to the Flat
model with Shortest Path Tree heuristics.

\subsection{Rooting\label{sub:Rooting}}

The PCSPG and SPG variants of the problem are undirected and unrooted
but the formalism introduced in Section \ref{sec:The-model} needs
to select a root node among the profitable or terminal nodes, for
the PCSPG and SPG variants respectively. It is always possible to
choose a random terminal for the SPG but there is no clear strategy
for the PCSPG. Moreover, the choice of a random terminal for the SPG
using the normal model could lead to a suboptimal solution for a limited
depth $D$. Here we propose one rooting procedure for each variant.

\subsubsection{SPG rooting}

Since the running time per iteration is $\Theta\left(D|E|\right)$,
it is convenient to select as root the terminal that allows a representation
with a small parameter $D$. A straightforward heuristics consists
then in finding the terminal for which the maximum hop distance to
other terminals is the smallest. This can be computed simply by an
application of Breadth-First Search for each root candidate, in time
proportional to $\left|K\right|\left|E\right|$ where $\left|K\right|$
is the number of terminals.

\subsubsection{PCSPG rooting}

The procedure reported in this section is the same used in \cite{biazzo_performance_2012}.
Add an extra root node $r$ and connect it to all profitable vertices
with extra edges of identical very large weight $\mu$. The optimal
PCSPG solution in this modified graph consists trivially in a single
node tree $\left\{ r\right\} $, since the addition of anything else
carries a cost $\mu$ that makes it unprofitable. Nevertheless, the
MS algorithm provides additional information besides the non-informative
optimal result. Consider the ``second'' optimum solution of the
problem. The root node $r$ will be connected to one (and only one)
vertex of the original graph, since adding additional edge costs equal
to $\mu$ will be cost-wise inconvenient. This vertex can be identified
by computing $j^{*}\in\arg\max_{j}H_{jr}\left(1\right)$. Now clearly,
nontrivial (unrooted) solutions of the original problem are in one
to one correspondence with $r$-rooted solutions of the modified problem
with only one neighbor of $r$ (and the difference in cost is simply
$\mu$). Unfortunately, the information contained in MS fields is
insufficient to reconstruct the full ``second'' optimal tree of
the modified graph so a second run of the MS algorithm in the original
graph, using $j^{*}$ as root is needed.

\subsection{Reinforcement \label{sub:Reinforcement}}

As explained in \cite{biazzo_performance_2012,bailly-bechet_prize-collecting_2009,bailly-bechet_finding_2011},
the reinforcement procedure speeds up and aids convergence of the
MS algorithm but adds an additional parameter $\gamma_{1}$. Smaller
values of $\gamma_{1}$ lead typically to better solutions at the
cost of longer convergence times. We adopted the following approach:
we start the MS algorithm with a large value of $\gamma_{1}\sim10^{-2}$.
Then we iteratively halve $\gamma_{1}$ and run again MS until $\gamma_{1}\sim10^{-5}$.
We can stop the loop in $\gamma_{1}$ if the energy gap between the
new solution and the old one is significantly small since reducing
again the parameter typically does generally not bring a significant
improvement.

\subsection{Depth \label{sub:Depth}}

The depth parameter $D$ is fundamental in the algorithm described
in this work. It unequivocally delimits the space of the solutions
since MS will provide $D$-bounded trees; furthermore the computational
time depends linearly on $D$. Choosing a small depth reduces the
running time but this can affect the quality of the solution: for
the SPG if $D$ is not sufficiently large the connection of all terminals
is impossible and in the PCSPG some profitable nodes cannot be reached.
Larger values of $D$ implies a larger solution space where the optimum
may be better. Thus we need to guarantee that all profitable vertices,
or terminals, can be connected within the tree and then let the algorithm
choose the optimal subgraph. 

The computation of the minimum value of the depth $D_{min}$ uses
again a Breadth-First Search procedure. We start searching from the
predefined root and we save the distances, in hops, of the shortest
paths between the root and any profitable vertex. We then choose as
starting value of $D$, the maximum value among all these lengths.
Notice that for the \textit{flat} representation $D_{min}$ can be
taken equal to the number of profitable vertices.

\subsection{Running schemes}

In the following we explain two operative procedures that will be
used for different experiments regarding the SPG and the PCSPG variants.
\begin{itemize}
\item \textit{D-increasing scheme.\label{D-increasing-scheme} }Once we
have determined the proper value of $D_{min}$ as in Subsection \prettyref{sub:Depth}
we apply MS and the heuristics following the reinforcement scheme
described in Subsection \prettyref{sub:Reinforcement}. The process
is repeated for increasing values of $D$ until the predefined running
time is over. 
\item \textit{D-bounded scheme. \label{D-bounded-scheme} }Here we run the
algorithm using a unique value of $D$. Simulations stop either because
the reinforcement scheme in Subsection \prettyref{sub:Reinforcement}
or the available time is over.
\end{itemize}

\section{\label{sec:Results-for-Scale-Grid}Results for Scale-free and Grid-like
graphs}

In this Section we report the performances of our new developments
for two classes of graphs which model very well real-world networks,
namely, the scale-free and the grid graphs.  We will show here that
the introduction of heuristics in Section \ref{sec:A-Belief-Propagation-Inspired}
guarantees feasible solutions after few iterations of MS even for
loopy graphs and we will underline the improvements carried by the
\textit{flat} model for particular instances. We show the results
of ``N'', ``J'' and ``W'' (combined to the normal or to the
$flat$ model) and we compare them to the results obtained by the
``old'' heuristic, ``O''; quantitatively we compute the percentage
gap between the energy given by algorithm $x$ and algorithm $y$
as:
\begin{equation}
Gap(x,\,y)=\frac{x-y}{y}\cdot100\label{eq:gap}
\end{equation}
where $x$ represents one of our ``new'' enhancements and $y$ the
``O'' algorithm; the more the gap is negative the more $x$ outperforms
$y$.

Experiments were run on a Multi-Core AMD opteron 2600Mhz server, where
most of the cases we use the \textit{D-increasing scheme} in Subsection
\ref{D-increasing-scheme} for a running time of 600 s; if different
schemes are used they will be precised in each section.

Most relevant results for single-instance problems are reported in
several tables in Appendix \ref{sec:Results-for-Scale} and \ref{sec:Results-for-Grid}
that have all the same structure. The first column contains the name
of the instances, the second one displays the best energy found by
the best algorithm that is reported in the third section. The remaining
columns list the running time needed by the best algorithm, the ``O''
result and the gap computed as in \eqref{eq:gap} between the best
energy and the ``O'' energy of the ``old'' algorithm. In some
specific frameworks we also propose averaged results over different
realizations of the same graph to confirm evidences suggested by single-instance
energies.

Instances names give clear information about graph properties. The
first letters identify the type of graph, i.e. can be either ``SF''
or ``G'' depending on we are dealing with a scale-free network or
a grid graph. They also contain the number of nodes and the number
of terminals that are followed by letters $n$ and $a$ respectively,
while a suffix ``-p'' is added for the PCSPG instances. Grid-graph
names also reveal the nodes layout on the graph. For instance, the
graph \textit{G\_100x100x2\_a10} is a 3d grid graph of size 100x100x2
that contains 20000 nodes, 10 of which are terminals and we aim to
solve a SPG.

\subsection{Performances of Max Sum against heuristics}

To give an example of the efficiency of the MS-guided heuristics we
create 5 grid graphs of size 10x10x10 containing terminals in the
range {[}10, 410{]} and 5 scale-free networks of $10^{3}$ nodes and
100 terminals with edges in the interval {[}2991, 10879{]} on which
we solve the SPG. In Table \ref{tab:OMSMST_G} and \ref{tab:OMSMST_SF}
we report the energies achieved by ``O'' algorithm, MS and the MST
without the reweighting scheme introduced by ``O''. Energies of
the MST algorithm are averaged over 10 realizations of each graph,
instead, for both ``O'' and MS we run the algorithms 10 times for
each of the 10 instances with different initial conditions and we
collect the best energies among the 10 initialitations. Then these
energies will be averaged over the 10 instances. The fraction of successes
over the 10$\times$10 attempts is reported in the left column of
``MS conv''. In the right column we count how many times MS converged
at least one time over the 10 initializations, and we normalize the
number of successes with respect to the number of instances per graph.

\begin{table}[H]
\begin{centering}
\begin{tabular}{c|c|c|r|c|c}
\textbf{Terminals} & \textbf{``O''} & \textbf{MS} & \multicolumn{1}{r}{\textbf{MS}} & \textbf{conv.} & \textbf{MST}\tabularnewline
\hline 
10 & 10.56 & 10.80  & 11/100 & 4/10 & 21.82\tabularnewline
\hline 
110 & 56.24 & 61.55  & 6/100 & 2/10 & 77.04\tabularnewline
\hline 
210 & 81.76 & 83.13  & 17/100 & 4/10 & 100.93\tabularnewline
\hline 
310 & 103.62 & 103.49  & 25/100 & 6/10 & 120.87\tabularnewline
\hline 
410 & 123.44 & 124.10  & 26/100 & 7/10 & 137.23\tabularnewline
\hline 
\end{tabular}
\par\end{centering}

\caption{Average energies for ``O'', MS, MST and MS convergences for grid
graphs 10x10x10 as a function of the number of terminals.\label{tab:OMSMST_G}}
\end{table}

\begin{table}[H]
\begin{centering}
\begin{tabular}{c|c|c|r|c|c}
\textbf{Edges} & \textbf{``O''} & \textbf{MS} & \multicolumn{1}{r}{\textbf{MS }} & \textbf{conv.} & \textbf{MST}\tabularnewline
\hline 
2991 & 34.09 & 33.89 & 30/100 & 4/10 & 40.72\tabularnewline
\hline 
4975 & 21.25 & 21.22 & 56/100 & 8/10 & 25.13\tabularnewline
\hline 
6951 & 16.53 & 16.92 & 59/100 & 7/10 & 18.52\tabularnewline
\hline 
8919 & 13.06 & 13.22 & 14/100 & 3/10 & 14.92\tabularnewline
\hline 
10879 & 10.85 & 10.76 & 30/100 & 3/10 & 12.49\tabularnewline
\hline 
\end{tabular}
\par\end{centering}

\caption{Average energies for ``O'', MS, MST and MS convergences for scale-free
graphs of 1000 nodes and 100 terminals as a function of the number
of edges. \label{tab:OMSMST_SF}}
\end{table}

Results suggest that MS very seldom converge on both families of networks
with an average fraction of success of 17/100 for grids and 38/100
for scale-free graphs. The heuristic ``O'' always outputs a solution
but sometimes is suboptimal in terms of energy as we can notice from
the comparison with the MS ones. Nevertheless these energies are far
below of the ones of the MST heuristic with original edge weights.

\subsection{Scale-free networks}

The first class of networks on which we run our algorithms is the
one of the scale-free graphs. Instances are generated by the Barabási-Albert
model also known as \textit{preferential attachment} scheme. Starting
from a $N_{0}$ nodes, new nodes are added, one at the time, to the
graph. Each new node is attached to $m$ different vertices with a
probability that is proportional to the connectivity of the existing
nodes. 

In our experiments the number of nodes of the graphs is $N=10^{4}$
for SPG instances, while the initial set of nodes $N_{0}$ has cardinality
$m$; all graphs have weights distributed according to the uniform
distribution in the interval {[}0, 1{]}.

\subsubsection{SPG results}

\begin{figure}[H]
\includegraphics[width=0.5\columnwidth]{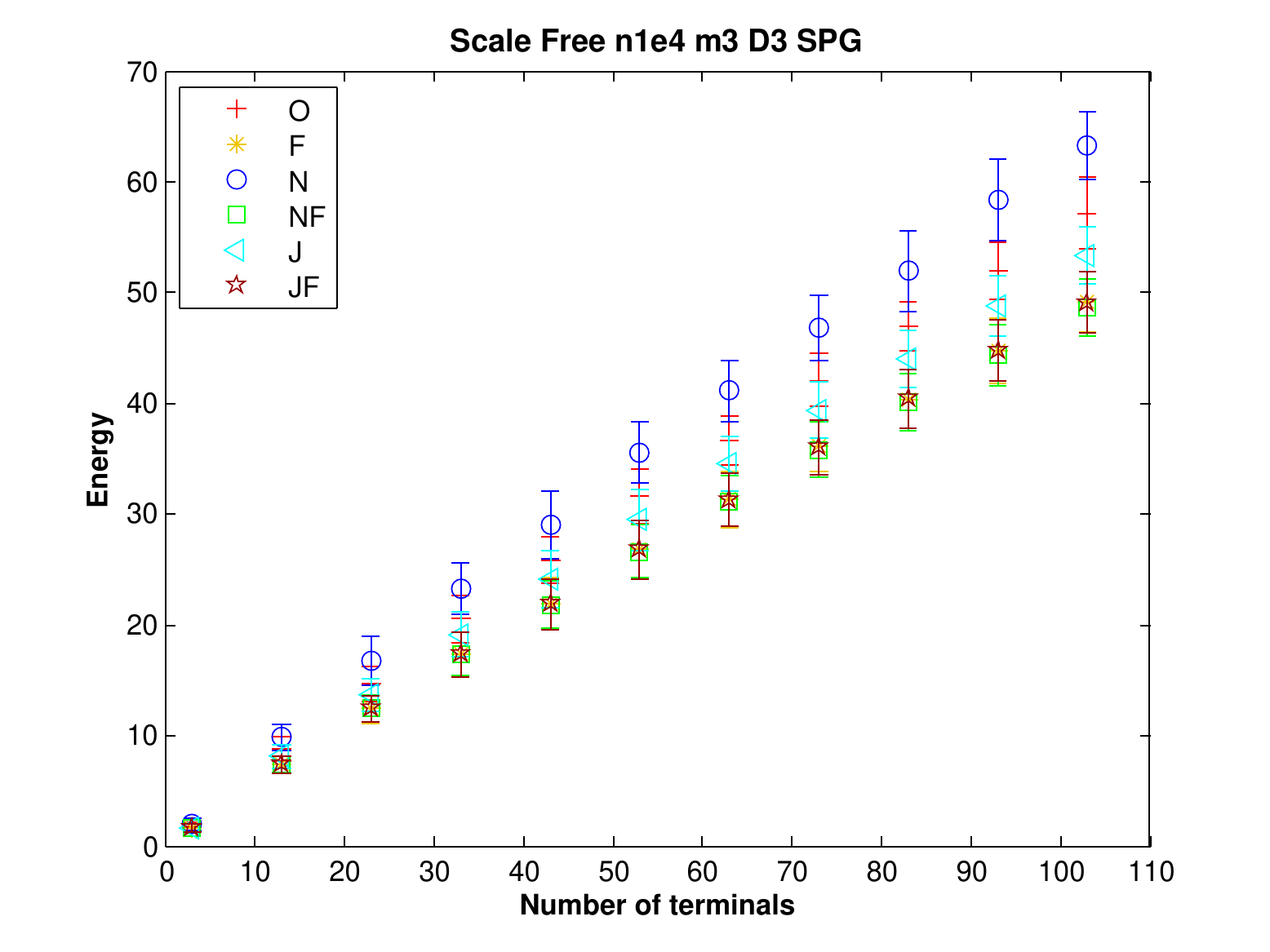}\includegraphics[width=0.5\columnwidth]{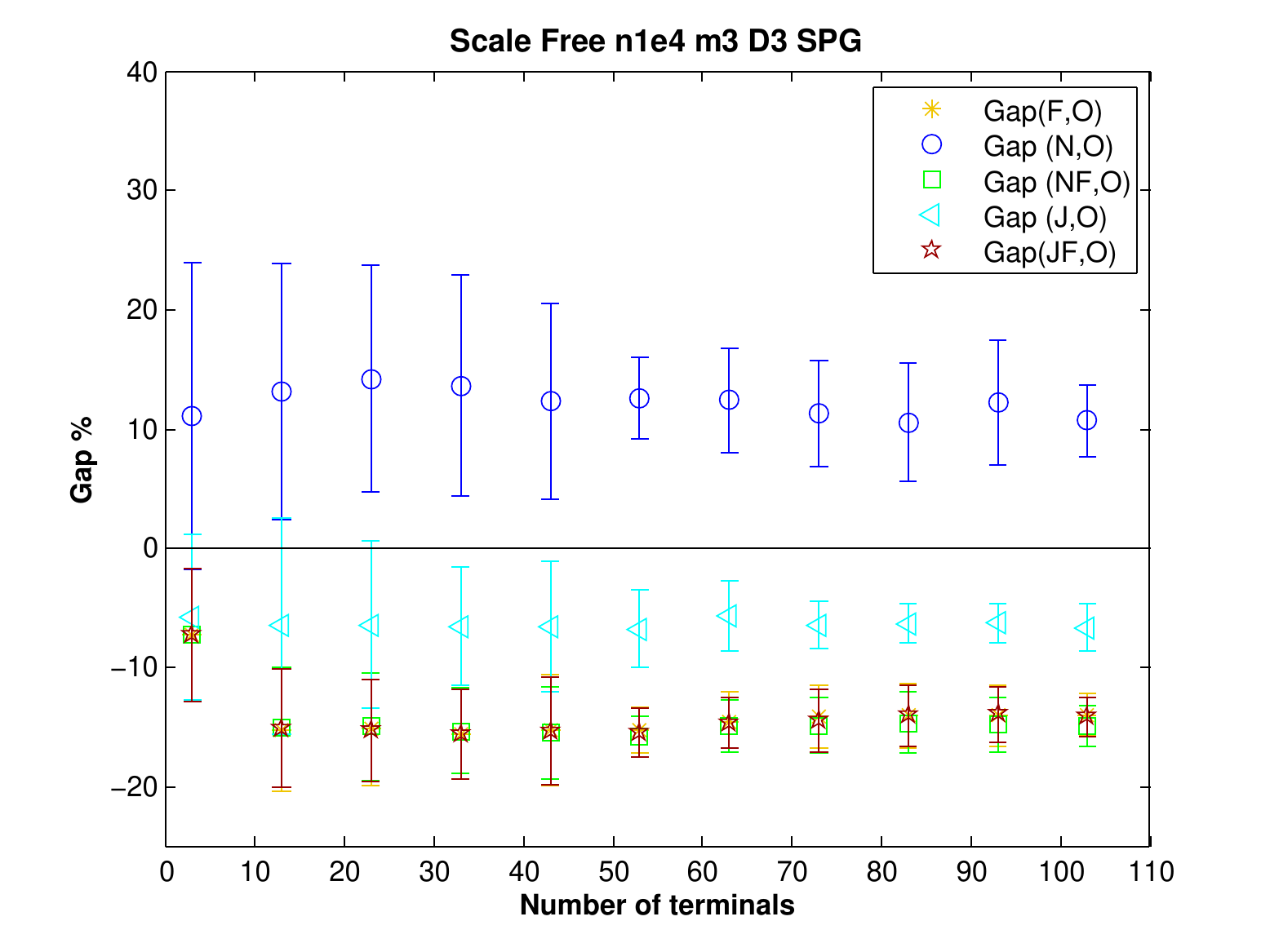}

\includegraphics[width=0.5\columnwidth]{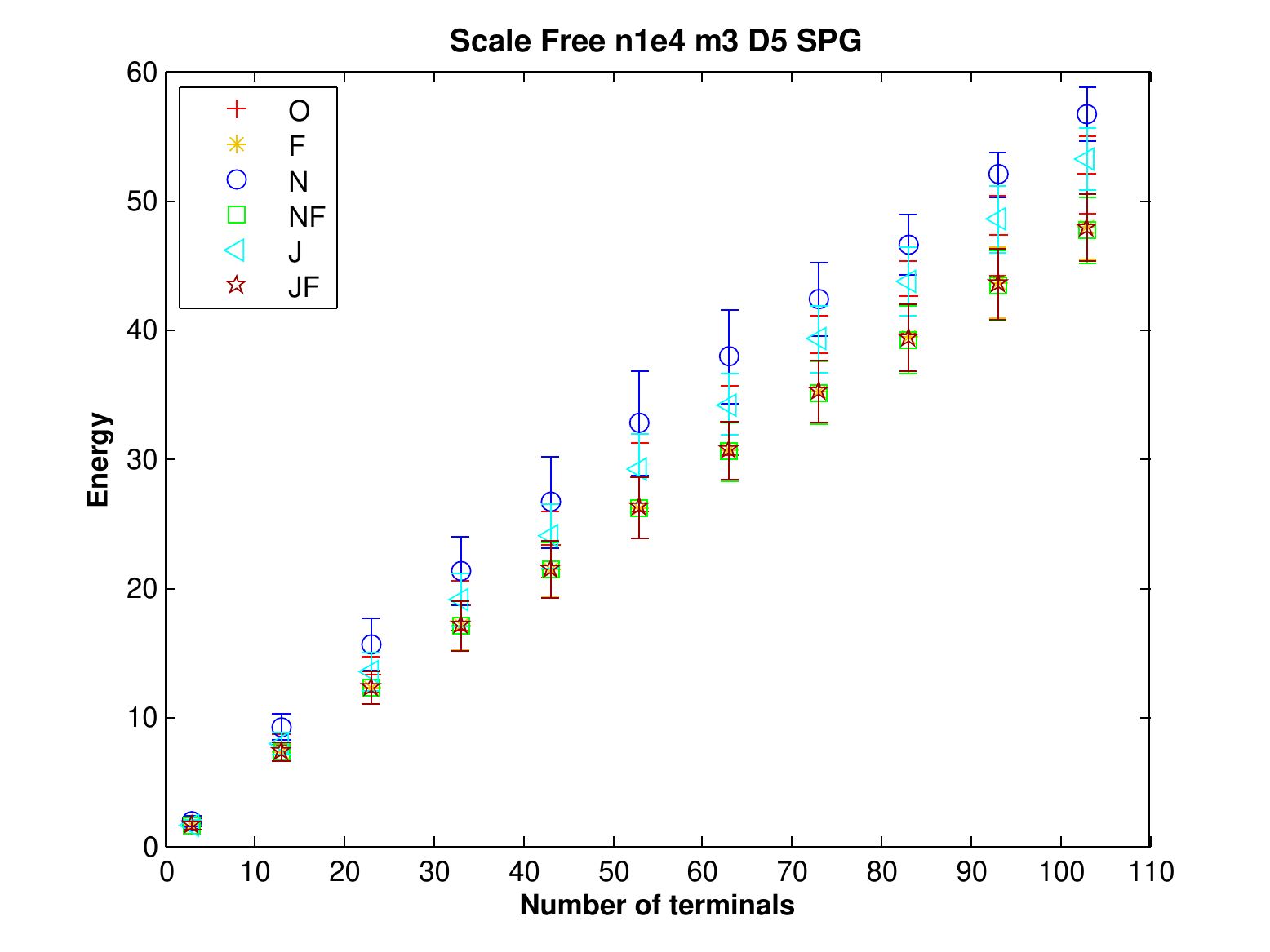}\includegraphics[width=0.5\columnwidth]{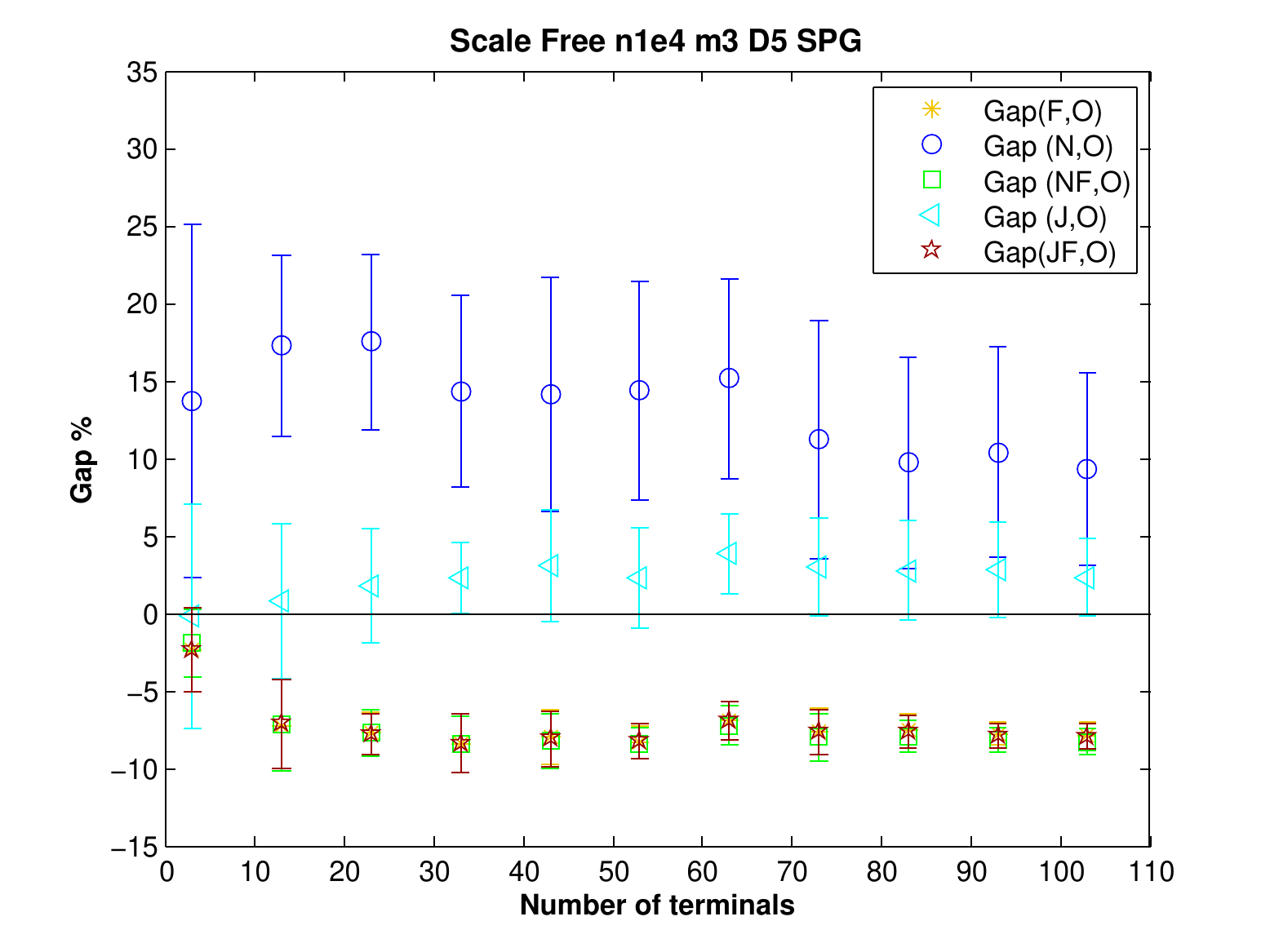}

\includegraphics[width=0.5\columnwidth]{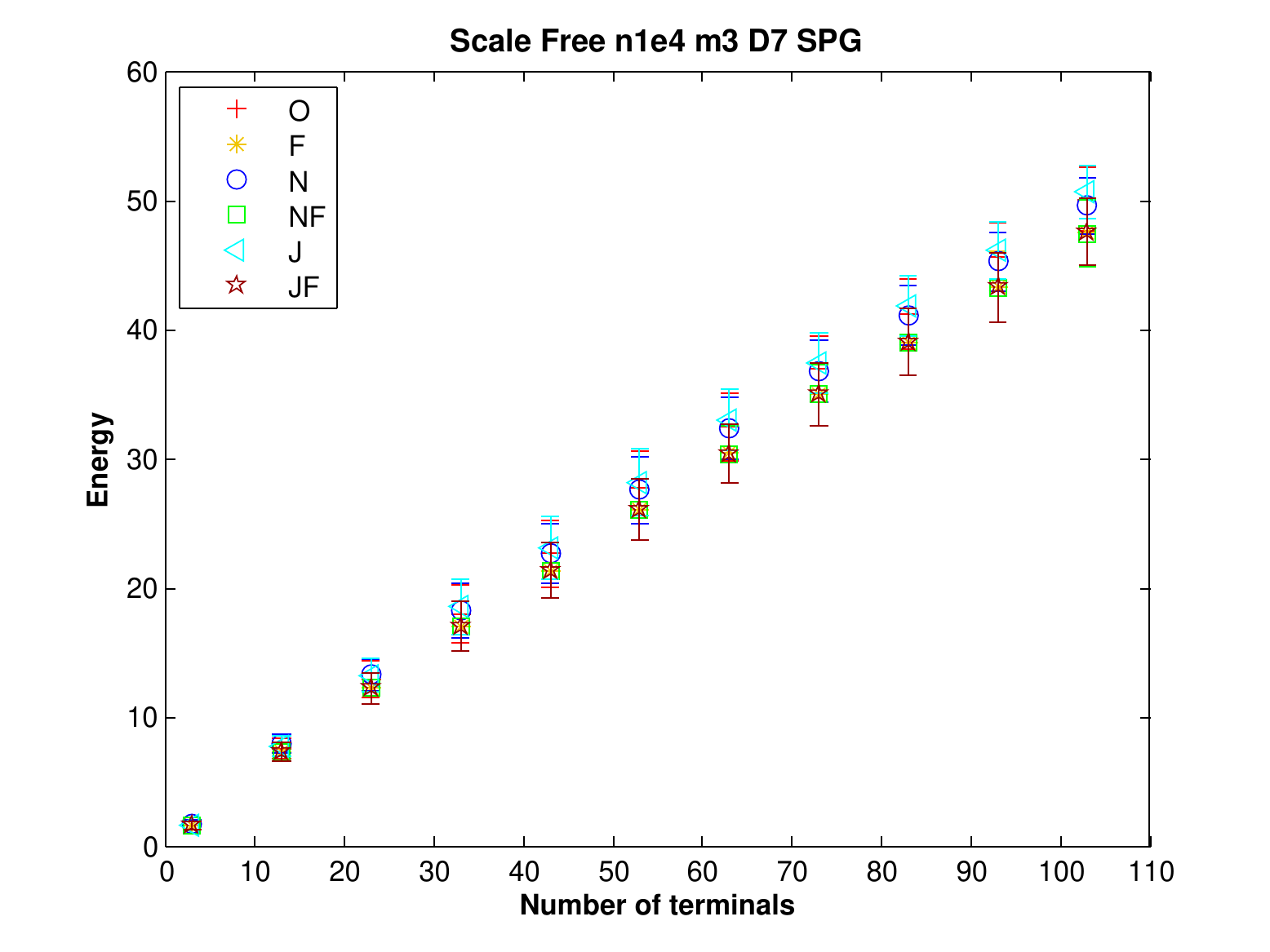}\includegraphics[width=0.5\columnwidth]{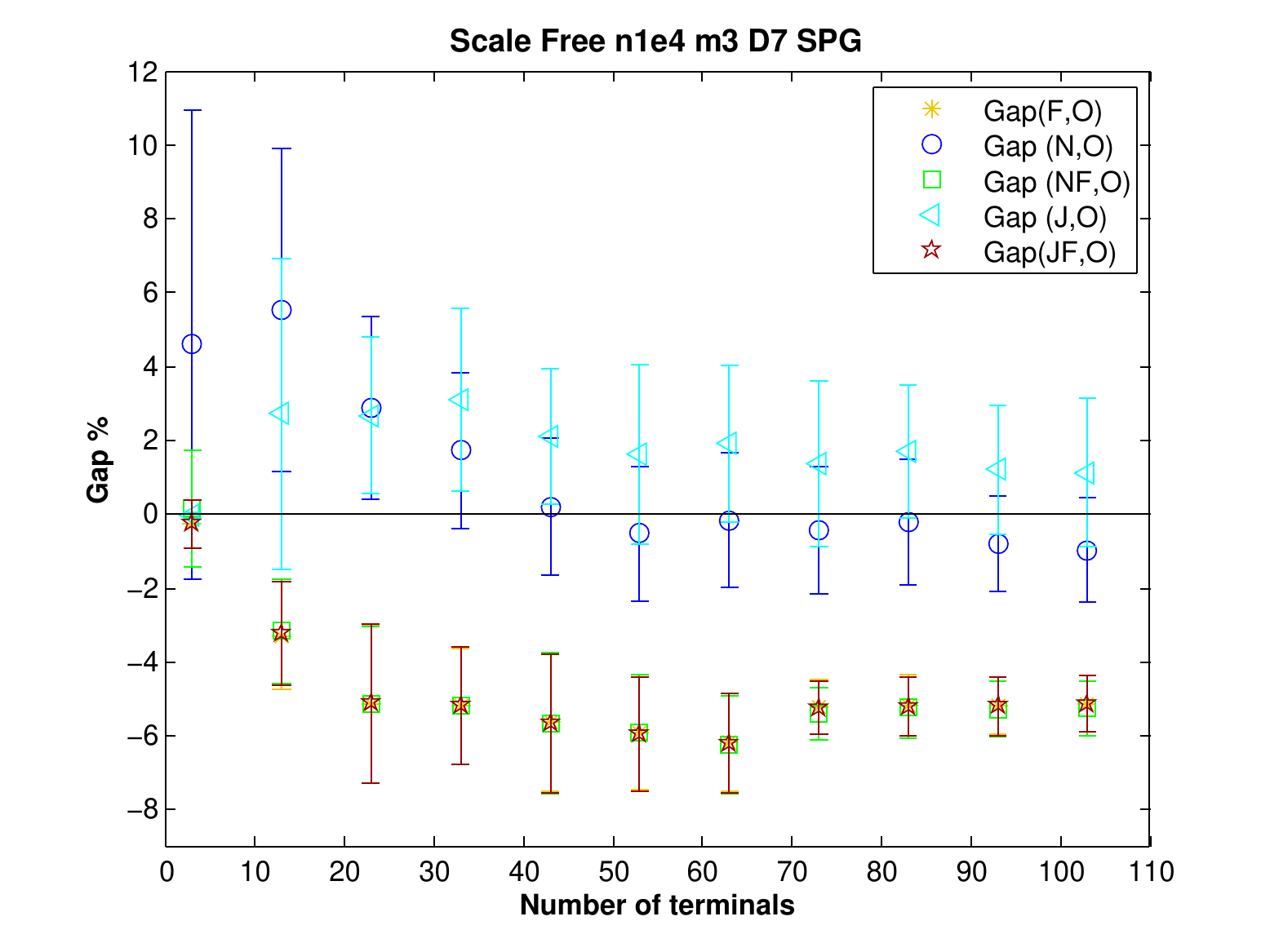}

\caption{Left: Energy of the solution for the SPG for scale-free graph as a
function of the number of terminals. Right: Energy gaps of the ``F'',
``N'', ``NF'', ``J'' and ``JF'' heuristics with respect to
``O''. \label{fig:SFEnGap} From up to the bottom the depth $D$
increases.}

\end{figure}

To underline the efficiency of the flat model we perform experiments
on sparse scale-free networks of $10^{4}$ nodes and $m=3$. We show
in Fig. \eqref{fig:SFEnGap} the energies and the gaps with respect
to ``O'' of all the other algorithms as a function of the number
of terminals for three values of the depth $D=\{3,5,7\}$ and unlimited
running time. For very small values of the depth (3 and 5) the ``N''
variant achieves gaps of the order of $10\%$ which is significantly
larger than the ones obtained by all \textit{flat}-model based heuristics
(``F'', ``NF'' and ``JF'') that on average are equal to $-15\%,\,-10\%$
and $-6$\% for increasing values of $D$. Only for $D=7$ the ``N''
variant reaches the ``O'' performances while the ``J'' heuristic
gives negative gaps only for $D=3$. 

In addition we show the performances for very dense graphs where the
parameter $m$ changes in the interval {[}10, 500{]} and we set $a=1000$.
As reported in Table \ref{tab:SFSPGa} the ``NF'' variant achieves,
for single-instance runs, the best performance as long as the number
of edges is not very large. As $m$ increases the ``N'' variant
outperforms all the other variants reaching gaps with respect to ``O''
that seem to increase, in absolute value, as we increment the number
of edges.

\subsection{Grid graphs}

\subsubsection{2d lattices}

For these simulations we create 20 graph of $10^{4}$ nodes lying
in a 100x100 square lattice and connected by edges whose weights distribution
is uniform in the interval {[}0 1{]}. The number of terminals $a$
varies from 10 to 1000 and their prizes, for PCSPG variant only, are
picked uniformly from the interval {[}0 15{]} to ensure a non-trivial
optimal solution different from a single-node tree, the root, (in
the case of very low prizes) and a spanning tree (in the opposite
case of very large prizes).

Regarding the SPG, best solutions for single-runs are obtained through
the ``F'' variant despite the gap with respect to the ``O'' algorithm
is on average equals to -0.2 \%. Besides, for graph containing a large
number of terminals, the most performing algorithm is the ``NF''.
As shown in Table \ref{tab:G2DPC} for the PCSPG the ``F'' and ``O''
variants achieve very close results except for graph \textit{G\_100x100\_a10-p}
where the ``W'' heuristic reaches a gap of -5 \%.

\subsubsection{3d grid-graphs}

Instances are 100x100x2 grid-graphs whose links have weights distributed
uniformly in {[}0, 1{]} and whose terminals, for PCSPG only, have
prizes in the range {[}0 3{]}. We investigate different regimes depending
on how many terminals are placed on the graph. 

Regarding the SPG variant, as illustrated in Table \ref{tab:G3DSPGfew},
the ``F'' variant more often gives the best solution in the case
of graphs with few terminals probably because here the ``depth''
for the ``F'' algorithm is the number of terminals which is much
smaller than the parameter $D_{min}$ computed as in Section \ref{sub:Depth}. 

\begin{figure}[H]
\includegraphics[width=0.5\columnwidth]{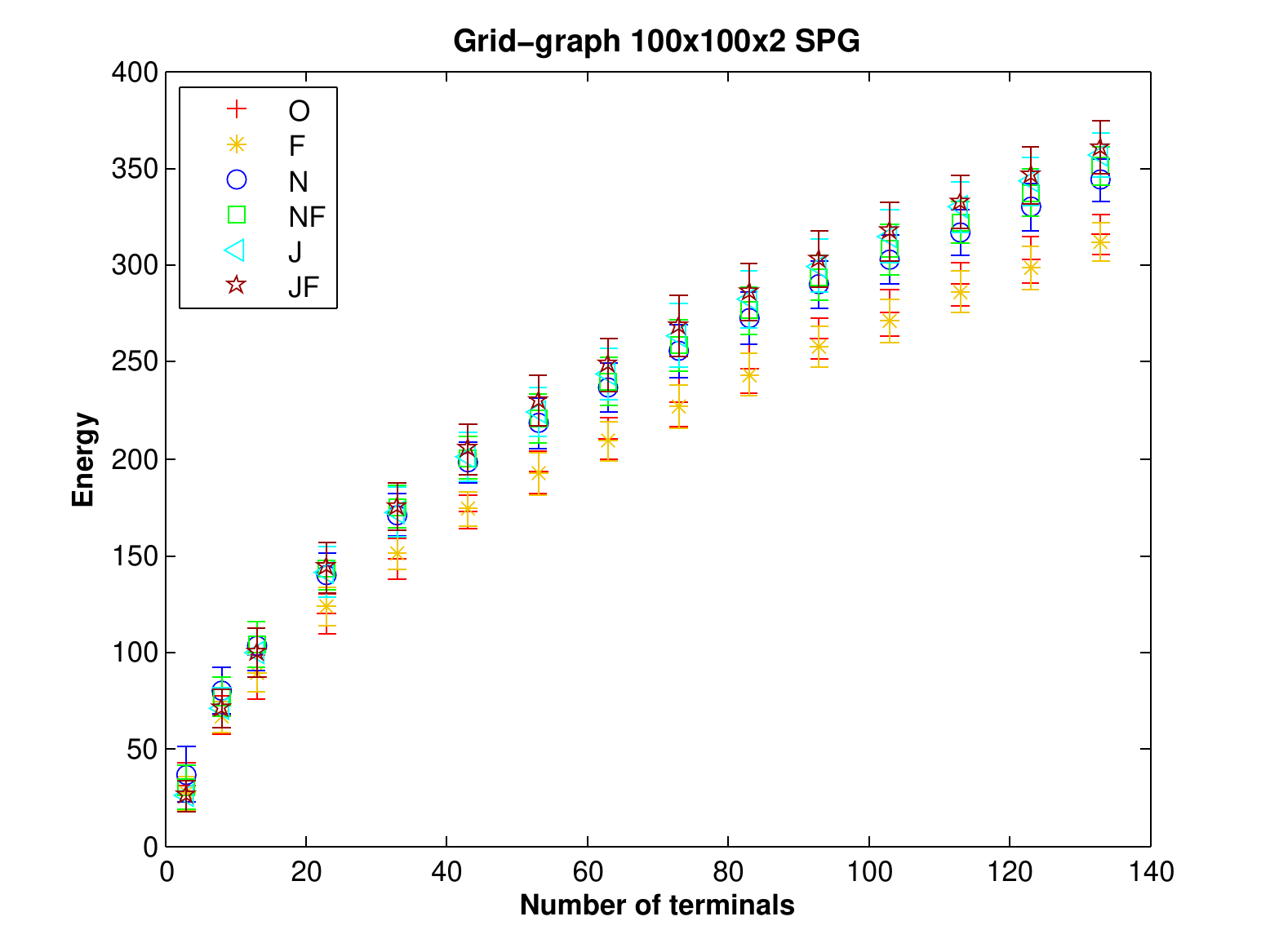}\includegraphics[width=0.5\columnwidth]{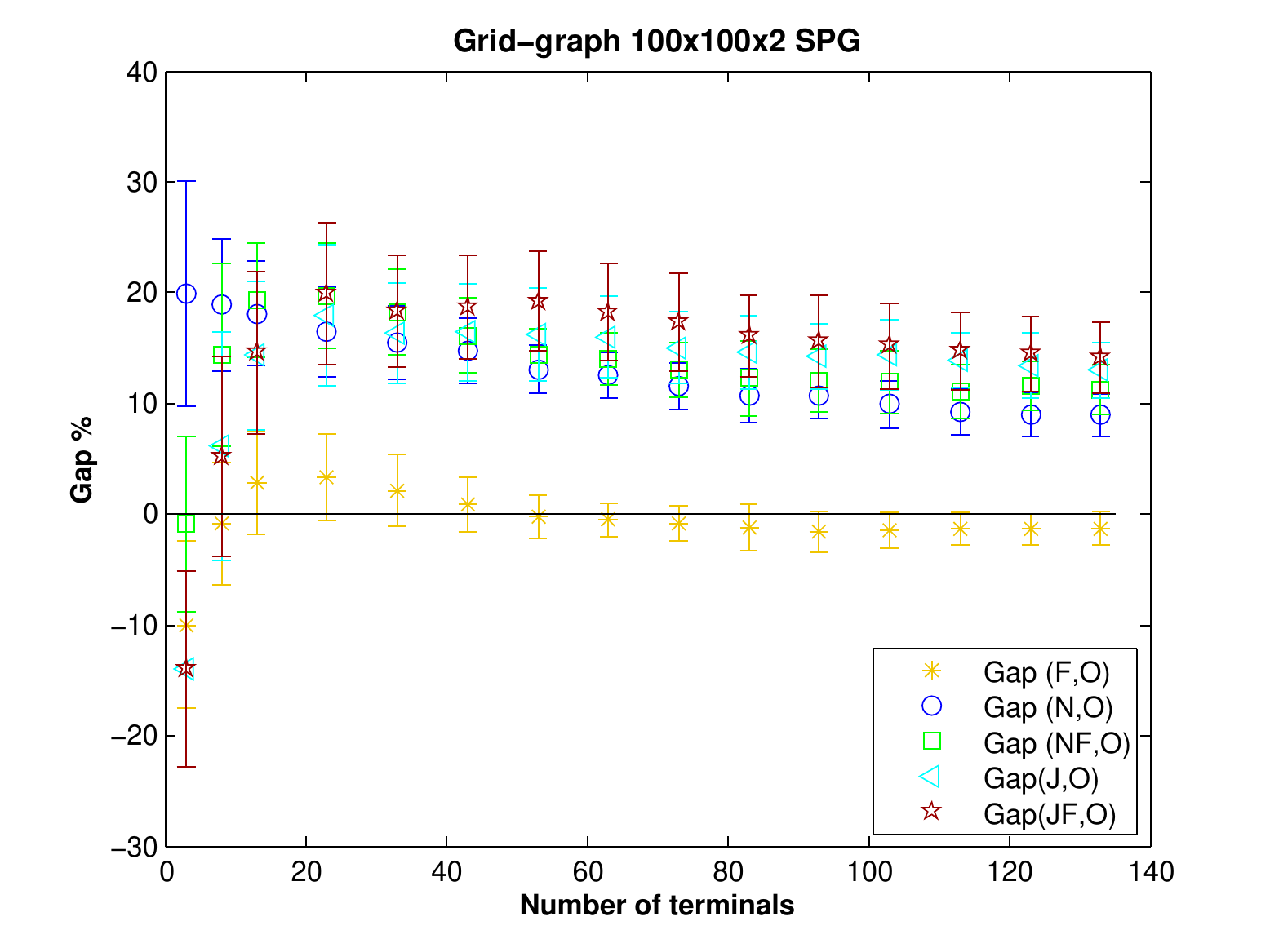}

\caption{Left: Energy of the solution for the SPG for a grid-graph 100x100x2
as a function of the number of terminals. Right: Energy gaps of the
``F'', ``N'', ``NF'', ``J'' and ``JF'' heuristics with respect
to ``O''. \label{fig:EnGapSPGgrid}}

\end{figure}

This statement is confirmed by the statistical measure of the energies
and the gaps as a function of the number of terminals that is shown
in Fig. \eqref{fig:EnGapSPGgrid}. The ``F'' variant, a part for
a small positive gap in the range {[}20, 40{]}, attains negative gaps
that reach -10\% for instances with few terminals and a constant -1\%
for all the range $a\in[80,140]$.

For the PCSPG we investigate how energies changes when the number
of terminals is in the range $[3000,10000]$. Single-instance results
suggest that the ``O'' variant outperforms all the other up to $a\sim4000$
where we observe that ``N'' and ``NF'' algorithms achieve best
results with a gap that decreases for increasing $a$ as reported
in Table \ref{tab:G3DPCmany}. 

\begin{figure}[H]
\includegraphics[width=0.5\columnwidth]{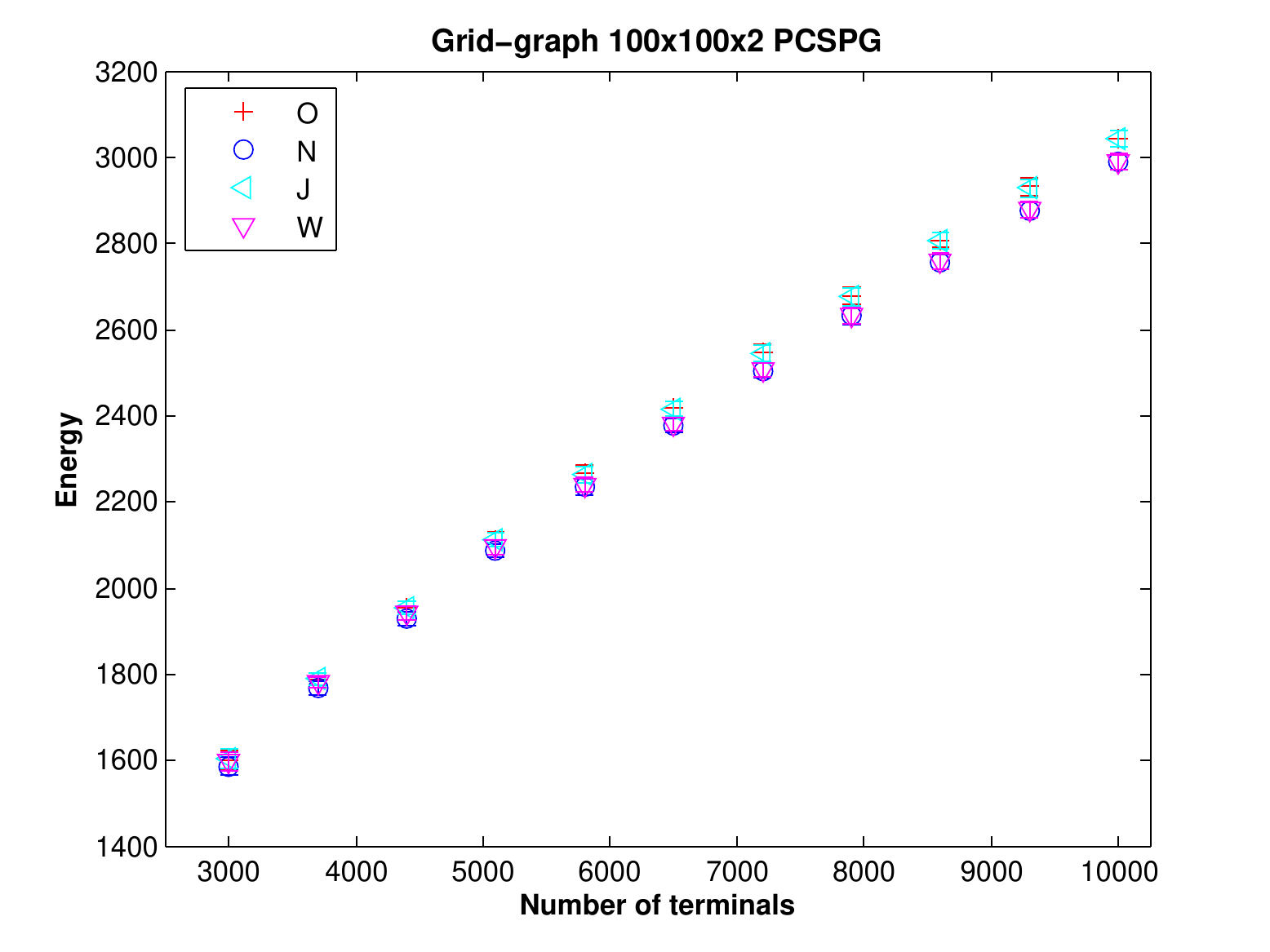}\includegraphics[width=0.5\columnwidth]{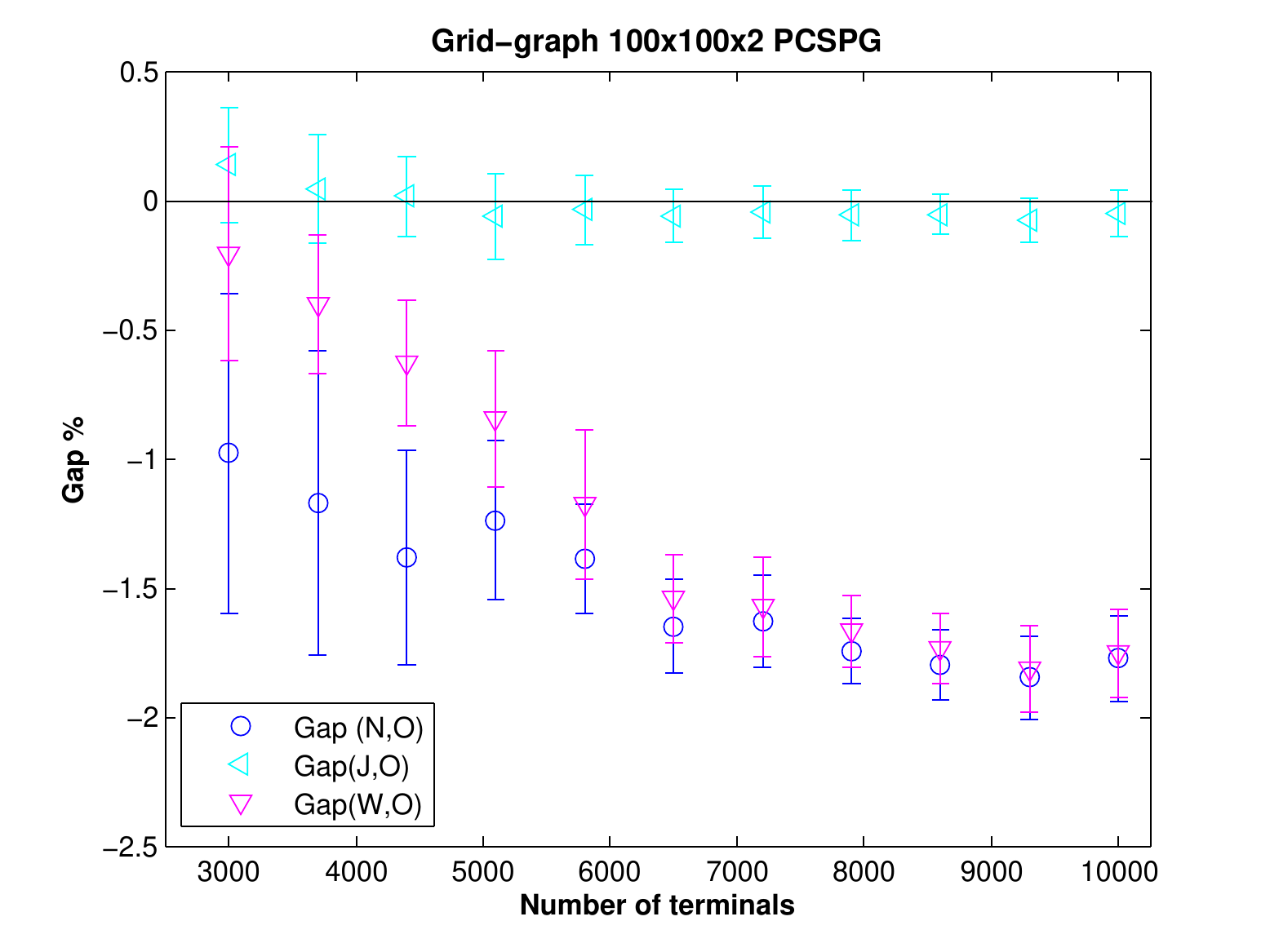}

\caption{Left: Energy of the solution for the PCSPG for a grid-graph 100x100x2
as a function of the number of terminals. Right: Energy gaps of the
``N'', ``J'' and ``W'' heuristics with respect to ``O''. \label{fig:EnGapPCgrid}}

\end{figure}

To fairly compare the results of the heuristics ``O'', ``N'',
``J'' and ``W'' in this regime, we perform several simulations
for each graph as $a$ varies in {[}3000, 10000{]} and we compare
the averaged outcomes in Fig. \eqref{fig:EnGapPCgrid}. While the
``J'' variant attains energies as similar to the ``O'' ones as
$a$ grows, the ``N'' and ``W'' heuristics increase their respective
gaps as $a$ increases.

\section{\label{sec:Results-for-DIMACS}Results of the DIMACS challenge and
StenLib instances}

Here we show the results of the MS algorithm combined to greedy heuristics
for the SPG and PCSPG benchmark chosen for DIMACS Challenge. These
instances have been selected to be particularly challenging (often
the optimal solutions are not known) and to be representative of the
whole set available here: \url{http://dimacs11.cs.princeton.edu/competition.html}.
Additionally we report our results for a set of hard instances, called
PUC, where most of the optimal solutions are still not known.

For both SPG and PCSPG variants we will need to compare outcomes provided
by different algorithms or to measure the improvements of the new
algorithm. According to the rules of the official competition, we
use as comparison metrics the Primal Bound (PB) that is the best solution
found in a fixed running time and the gap, computed as in \eqref{eq:gap},
between algorithms $x$ and $y$ that will be specified in the following
sections. Absolute values of the PB and the respectively gaps are
collected in tables which are all displayed in Appendix \ref{sec:Dimacs=000026StenLib}. Notice that official results are often rounded to the first decimal place and this may slightly affect the accuracy of the gaps.

As for the scale-free and grid graph, simulations were run on a Multi-Core
AMD opteron 2600Mhz server, which is slightly slower than the cluster
on which the challenge simulations were performed. Nevertheless, we
compared the results obtained in the challenge by our submitted code
on our local server with the same time limit, and the gaps of the
primal bounds between ``O'' and ``polito'' are on average $\pm0.3$\%
for all instances, well inside the confidence interval for these simulations.
Thus we are sufficiently confident that new results can be also compared
fairly with those of the challenge.

\subsection{SPG results}

\subsubsection{Preprocessing}

Regarding the SPG problem, a common approach is to apply some reduction
or preprocessing techniques to instances. Preprocessing may significantly
modify the original graph but in a way that an optimal solution of
the reduced graph can be easily mapped into the minimum Steiner Tree
of the original problem. The advantages of preprocessing consist in
a speed up of the convergence and, often, a clear improvement of the
solutions. In this work we use the preprocessing feature of the nice
package Bossa, \url{http://www.cs.princeton.edu/~rwerneck/bossa/}. 

We use the label ``pre'' for the results in which instances are
preprocessed before the application of the algorithm. In all cases
in which it was measured, the running time \emph{include} the preprocessing
time.

\subsubsection{Results for the D-increasing scheme}

In the following we present the best Primal Bounds among all variants
of the algorithm labelled according to Section \ref{sub:Labelling}.
Experiments are performed as in the \textit{D-increasing scheme} in
Section\ref{D-increasing-scheme} where the running time is set to
be of 7200 s.

Results are displayed in Table \ref{PB_SPG_table}. The second column
reveals the best Primal Bound found using our algorithms while the
third one displays which model and/or heuristics reach such results.
In the third section we report the running time of our fastest performer
which is specified in brackets. The last two sections are devoted
to the comparison between the best new performance and the ``polito''
PB. 

Generally we improved our old ``polito'' results with some significant
gaps of -5.87 \% and -4.11 \% for \textit{word666} and \textit{es10000fst01}
instances respectively. Furthermore we notice that, globally, all
``N-like'' options outperforms the other variants.

In Table \ref{PB_other} we compare our new results to the best results
of the competition. For each instance we show our PB, the best PB
found in the challenge, where in brackets we report the performer
that attained such result, and the last column contains the gaps between
the two energies computed according to \eqref{eq:gap}. Despite we
are quite far from the best known bound for some instances (for example
\textit{alut2625}, \textit{lin36} and \textit{lin37} whose gaps are
higher than 10 \%), we generally approached the performance of the
best challengers of the competition. Moreover, we further improved
the best known energies of some solutions, like for the \textit{i640-341}
and the \textit{``cc12}-like'' instances; such primal bounds are
reported in bold letters. 

Motivated by the performances of the ``N-like'' algorithm and by
the improvements obtained on the \textit{``cc12}-like'' instances,
we run ``Npre'' on the entire set of the PUC instances to which
these graphs belong. Results shown if Table \ref{PUC} and \ref{PUC2}
reveals that we achieve optimal performances since all gaps between
Npre and the best known energies are smaller then 0.80 \%. Moreover,
we reach new best known bounds for \textit{cc10-2p}, \textit{cc11-2p}
and \textit{hc11p} instances.

\subsubsection{Results for the D-bounded scheme \label{flat_vs_normal}}

To underline the differences between the branching and the \textit{flat}
model, we run again the algorithm using the \textit{D-bounded scheme}
in Section \vref{D-bounded-scheme} imposing the depth equals to the
number of terminals. The running time is set to 1200 s for these experiments.

\begin{figure}[H]
\begin{centering}
\includegraphics[width=1\columnwidth]{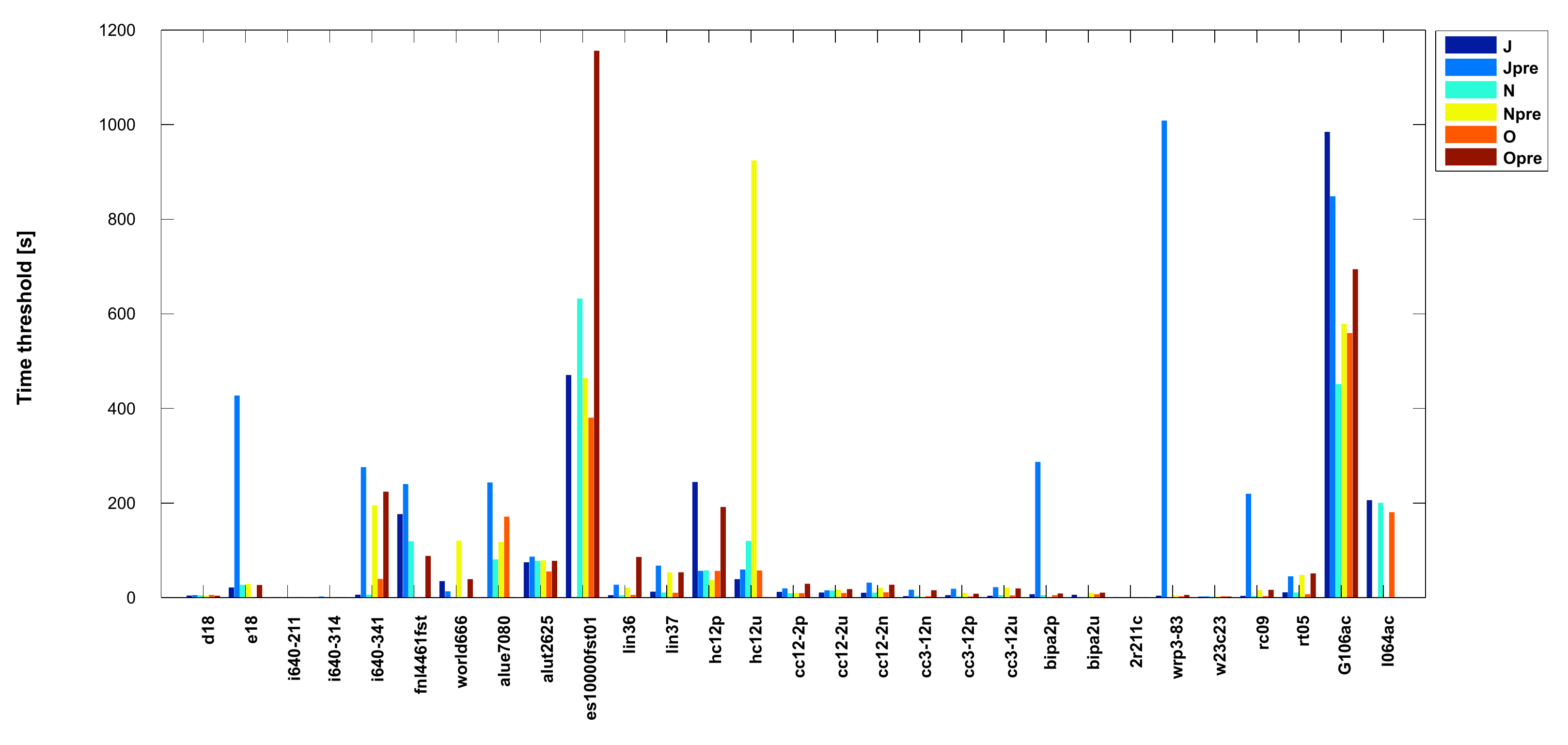}
\par\end{centering}

\caption{Time threshold Flat vs Branching model\label{fig:Time-threshold-Flat}}
\end{figure}

For each heuristic, ``N'', ``O'' and ``J'', we compute the time
interval for which the same heuristics combined with the \textit{flat}
model, ``NF'', ``F'' and ``JF'', provides a better solution
then the normal model; the same experiment is performed for the pre-processed
instances. In Fig. \eqref{fig:Time-threshold-Flat} we report such
time thresholds. 

For several instances the time threshold is equal to zero, meaning
that the branching representation is always better than the \textit{flat}
one. A threshold of the order of hundreds of seconds appears for several
instances while for \textit{es1000fst101}, a rectilinear graph, and
the \textit{G106ac}, a ``Vienna class'' graph, the time interval
in which the \textit{flat} model (combined to almost all the heuristics)
provides the best solution is significantly large.

\subsection{PCSPG results}

Here we show the performance of our new enhancements ``F'', ``J'',
``N'' and ``W'' for the PCSPG variant. As for the SPG case we
first collect our best primal bounds and we compare them to the ``polito''
results of the challenge. Afterwards, we show the measure of the gap
with respect to the three best algorithms of the competition, ``KTS'',
``staynerd'' and ``mozartballs''. We perform simulations using
the operative scheme in Section \ref{D-increasing-scheme} for a running
time of 7200 s. 

As plotted in Table \ref{PB_PC_best} we find that the best performer
is ``F'' that outputs the best PB for 27 instances over 32 but the
more significant improvements are given by ``W'' for the \textit{K400-7},
\textit{hands04} and \textit{handbd13} instances with a gap of -7
\%, -10\% and -27 \% respectively; the fastest performer instead is
the ``N'' variant.

In Table \ref{PB_PC_chal} we show the results of our best performers
compared to ``KTS'', ``staynerd'', ``mozartballs'' and ``polito''
algorithms of the competition. Energies are now well comparable to
the best known PB with an average gap of 0.1 \% for 29 instances.
Moreover, we achieve the best known energy for \textit{hc10p}, \textit{hc11u},
\textit{hc12u} and \textit{cc12-2nu} by our new algorithm; these results
are reported in bold letters in Table \ref{PB_PC_chal}. Note that
additionally, for the instance \textit{hc12p}, the best known energy
is the one obtained by ``polito'', i.e. the ``O'' variant.

\subsection{Non-reweighting scheme results}

In order to underline the improvements of the MS-based reweighting
scheme for these hard instances, we compare them to the PB provided
by the heuristics (pruned Minimum Spanning Tree, pruned Shortest Path
Tree and Goemans-Williamson heuristics) for the SPG and PCSPG instances
with original weights on edges. As reported in Table \ref{Heur_MS}
and \ref{Heur_MS2} energies in columns ``MST'', ``SPT'' and ``GW''
(for Prize-collecting only) are far from being comparable to the performances
of the re-weighted scheme, labelled as ``MS reweighting'', and thus
to the state-of-the-art algorithms of the challenge.

\section{\label{sec:Conclusion}Conclusion}

In this work, we presented several improvements for a message-passing
approach to several variants of the Steiner Tree Problem on graphs.
A first improvement is the incorporation of heuristics that are able
to transform partial information coming from an intermediate state
of the messages before convergence into feasible solutions. This may
be useful when the computation time at disposal is short, but it is
of particular importance because it forces the algorithm to output
solutions even in cases in which the tree-like approximation is inaccurate
and reinforced Max-Sum equations do not converge. This resulted in
the variant that we called ``O'' in this manuscript, that participated
in the 2014 DIMACS Challenge with encouraging results. We give a detailed
report on the outcome of the Challenge.

With respect to the results in the Challenge, we report here several
further improvements, including two different heuristics, the ``W'',
``N'' and ``J'' variants and their combinations. Many strategies
have been explored and reported here, with the outcome that the ``N''
variant is the best overall, but some of the other variants give advantages
for some instance types. During this development we derived an ``edge
variables'' formulation, that allows to cope with a modified \textit{flat
}model that removes one impediment of past approaches of this type;
namely the need of a large maximum hop-distance $D$. This results
in the ``F'' variant of the algorithm, that give advantages in particular
on structured networks including meshes and scale-free graphs. Moreover
the ``edge variables'' formulation presented here is also in principle
able to accommodate other constraints, such as degree ones, in a simple
way. 
\begin{acknowledgments}
AB acknowledges support by Fondazione CRT under the initiative ``La Ricerca dei Talenti''. 
The authors acknowledge ERC grant No. 267915 for financial support in the participation to the DIMACS challenge. 
\end{acknowledgments}

\pagebreak{}

\appendix

\section{Results for Scale free Graphs\label{sec:Results-for-Scale}}

\begin{table}[H]
\noindent \begin{centering}
\begin{tabular}{l|.l.|..}
  \multicolumn{1}{c|}{\textbf{Instance}} & \multicolumn{1}{c}{\textbf{PB}} &\multicolumn{1}{c}{ \textbf{Algo} }&\multicolumn{1}{c|}{ \textbf{Time {[}s{]}}} &\multicolumn{1}{c}{ \textbf{``O'' PB}} &\multicolumn{1}{c}{ \textbf{Gap \%}}\tabularnewline
\hline 
SF\_n1e4\_m10\_a1e3 & 115.92 & N  & 595.97  & 116.86 & -0.80\tabularnewline
\hline 
SF\_n1e4\_m36\_a1e3 & 33.54 & NF   & 266.24   & 36.24 & -7.45\tabularnewline
\hline 
SF\_n1e4\_m62\_a1e3 & 19.28 & NF   & 381.05   & 20.64 & -6.59\tabularnewline
\hline 
SF\_n1e4\_m87\_a1e3 & 14.21 & NF   & 492.69   & 15.18 & -6.36\tabularnewline
\hline 
SF\_n1e4\_m113\_a1e3 & 11.37 & NF   & 443.22   & 12.67 & -10.26\tabularnewline
\hline 
SF\_n1e4\_m139\_a1e3 & 9.14 & NF   & 602.81   & 10.39 & -12.00\tabularnewline
\hline 
SF\_n1e4\_m165\_a1e3 & 8.89 & N  & 487.00   & 8.92 & -0.41\tabularnewline
\hline 
SF\_n1e4\_m191\_a1e3 & 7.24 & N  & 29.51   & 7.33 & -1.24\tabularnewline
\hline 
SF\_n1e4\_m216\_a1e3 & 6.80 & N  & 325.07   & 6.90 & -1.44\tabularnewline
\hline 
SF\_n1e4\_m242\_a1e3 & 5.84 & N  & 40.44   & 5.92 & -1.48\tabularnewline
\hline 
SF\_n1e4\_m268\_a1e3 & 5.67 & N  & 454.76   & 5.75 & -1.40\tabularnewline
\hline 
SF\_n1e4\_m294\_a1e3 & 5.05 & N  & 60.07   & 5.14 & -1.79\tabularnewline
\hline 
SF\_n1e4\_m319\_a1e3 & 4.58 & N  & 55.55   & 4.76 & -3.83\tabularnewline
\hline 
SF\_n1e4\_m345\_a1e3 & 4.07 & N  & 46.22   & 4.19 & -3.05\tabularnewline
\hline 
SF\_n1e4\_m371\_a1e3 & 4.05 & N  & 49.51   & 4.29 & -5.63\tabularnewline
\hline 
SF\_n1e4\_m397\_a1e3 & 3.86 & N  & 76.40   & 4.05 & -4.69\tabularnewline
\hline 
SF\_n1e4\_m423\_a1e3 & 3.44 & N  & 106.89   & 3.74 & -7.92\tabularnewline
\hline 
SF\_n1e4\_m448\_a1e3 & 3.29 & N  & 80.44   & 3.62 & -9.01\tabularnewline
\hline 
SF\_n1e4\_m474\_a1e3 & 3.14 & N  & 96.80   & 3.42 & -8.00\tabularnewline
\hline 
SF\_n1e4\_m500\_a1e3 & 3.10 & N  & 104.78   & 3.35 & -7.37\tabularnewline
\hline 
\end{tabular}
\par\end{centering}

\caption{Best Primal Bounds for SPG on Scale Free networks, fixed \textit{a\label{tab:SFSPGa}}}
\end{table}

\pagebreak{}

\section{Results for Grid Graphs\label{sec:Results-for-Grid}}

\begin{table}[H]
\noindent \begin{centering}
\begin{tabular}{l|.l.|..}
\multicolumn{1}{c|}{\textbf{Instance}} & \multicolumn{1}{c}{\textbf{PB}} & \multicolumn{1}{c}{\textbf{Algo}} & \multicolumn{1}{c|}{\textbf{Time {[}s{]}}} & \multicolumn{1}{c}{\textbf{``O'' PB}} & \multicolumn{1}{c}{\textbf{Gap \%}}\tabularnewline
\hline 
G\_100x100\_a10-p & 50.36 & W  & 13.977 & 53.18 & -5.29\tabularnewline
\hline 
G\_100x100\_a62-p & 215.10 & O   & 25.072 & 215.10 & -\tabularnewline
\hline 
G\_100x100\_a114-p & 305.92 & O   & 8.24   & 305.92 & -\tabularnewline
\hline 
G\_100x100\_a166-p & 363.03 & F   & 20.32   & 369.46 & -1.74\tabularnewline
\hline 
G\_100x100\_a218-p & 421.11 & F   & 206.27  & 427.23 & -1.43\tabularnewline
\hline 
G\_100x100\_a271-p & 477.41 & O   & 13.62   & 477.41 & -\tabularnewline
\hline 
G\_100x100\_a323-p & 532.13 & F   & 34.91   & 538.51 & -1.18\tabularnewline
\hline 
G\_100x100\_a375-p & 554.55 & O   & 98.42   & 554.55 & -\tabularnewline
\hline 
G\_100x100\_a427-p & 581.97 & F   & 168.98  & 586.18 & -0.72\tabularnewline
\hline 
G\_100x100\_a479-p & 624.02 & O   & 16.64   & 624.02 & -\tabularnewline
\hline 
G\_100x100\_a531-p & 654.29 & O   & 16.22   & 654.29 & -\tabularnewline
\hline 
G\_100x100\_a583-p & 673.97 & F   & 73.10   & 675.27 & -0.19\tabularnewline
\hline 
G\_100x100\_a635-p & 725.31 & O   & 12.44   & 725.31 & -\tabularnewline
\hline 
G\_100x100\_a687-p & 742.67 & O   & 13.62   & 742.67 & -\tabularnewline
\hline 
G\_100x100\_a739-p & 765.91 & F   & 247.15  & 770.46 & -0.59\tabularnewline
\hline 
G\_100x100\_a792-p & 788.36 & F   & 261.86  & 795.58 & -0.91\tabularnewline
\hline 
G\_100x100\_a844-p & 817.39 & O   & 14.45   & 817.39 & -\tabularnewline
\hline 
G\_100x100\_a896-p & 854.57 & O   & 13.51   & 854.57 & -\tabularnewline
\hline 
G\_100x100\_a948-p & 872.88 & O   & 11.95   & 872.88 & -\tabularnewline
\hline 
G\_100x100\_a1000-p & 890.94 & O   & 10.96   & 890.94 & -\tabularnewline
\hline 
\end{tabular}
\par\end{centering}

\caption{Primal Bounds for the PCSPG on 2d grid graphs\label{tab:G2DPC}}
\end{table}

\pagebreak{}

\begin{table}[H]
\noindent \begin{centering}
\begin{tabular}{l|.l.|..}
\multicolumn{1}{c|}{\textbf{Instance}} & \multicolumn{1}{c}{\textbf{PB}} & \multicolumn{1}{c}{\textbf{Algo}} & \multicolumn{1}{c|}{\textbf{Time {[}s{]}}} & \multicolumn{1}{c}{\textbf{``O'' PB}} & \multicolumn{1}{c}{\textbf{Gap \%}}\tabularnewline
\hline 
G\_100x100x2\_a10 & 78.07 & O   & 51.142 & 78.07 & -\tabularnewline
\hline 
G\_100x100x2\_a62 & 203.13 & O   & 14.114 & 203.13 & -\tabularnewline
\hline 
G\_100x100x2\_a114 & 286.96 & F   & 26.15   & 290.06 & -1.07\tabularnewline
\hline 
G\_100x100x2\_a166 & 358.52 & F   & 25.08   & 362.57 & -1.12\tabularnewline
\hline 
G\_100x100x2\_a218 & 424.50 & F   & 31.80   & 429.52 & -1.17\tabularnewline
\hline 
G\_100x100x2\_a271 & 468.21 & O   & 27.04   & 468.21 & -\tabularnewline
\hline 
G\_100x100x2\_a323 & 495.39 & F   & 57.39   & 503.96 & -1.70\tabularnewline
\hline 
G\_100x100x2\_a375 & 552.07 & F   & 32.98   & 556.24 & -0.75\tabularnewline
\hline 
G\_100x100x2\_a427 & 578.32 & F   & 39.57   & 586.57 & -1.41\tabularnewline
\hline 
G\_100x100x2\_a479 & 608.99 & F   & 55.25   & 609.44 & -0.07\tabularnewline
\hline 
G\_100x100x2\_a531 & 635.29 & F   & 43.12   & 644.63 & -1.45\tabularnewline
\hline 
G\_100x100x2\_a583 & 662.13 & F   & 44.29   & 665.76 & -0.55\tabularnewline
\hline 
G\_100x100x2\_a635 & 692.11 & F   & 47.99   & 697.51 & -0.77\tabularnewline
\hline 
G\_100x100x2\_a687 & 716.63 & F   & 50.44   & 719.79 & -0.44\tabularnewline
\hline 
G\_100x100x2\_a739 & 755.78 & O   & 30.24   & 755.78 & -\tabularnewline
\hline 
G\_100x100x2\_a792 & 780.45 & F   & 56.66   & 781.62 & -0.15\tabularnewline
\hline 
G\_100x100x2\_a844 & 797.13 & F   & 38.55   & 804.08 & -0.86\tabularnewline
\hline 
G\_100x100x2\_a896 & 822.84 & F   & 42.10   & 823.93 & -0.13\tabularnewline
\hline 
G\_100x100x2\_a948 & 843.94 & F   & 49.80   & 847.32 & -0.40\tabularnewline
\hline 
\end{tabular}
\par\end{centering}

\caption{Primal Bounds for the SPG on 3d grid graphs with few terminals\label{tab:G3DSPGfew}}
\end{table}

\pagebreak{}

\begin{table}[H]
\noindent \begin{centering}
\begin{tabular}{l|.l.|..}
\multicolumn{1}{c|}{\textbf{Instance}} & \multicolumn{1}{c}{\textbf{PB}} & \multicolumn{1}{c}{\textbf{Algo}} & \multicolumn{1}{c|}{\textbf{Time {[}s{]}}} & \multicolumn{1}{c}{\textbf{``O'' PB}} & \multicolumn{1}{c}{\textbf{Gap \%}}\tabularnewline
\hline 
G\_100x100x2\_a1000-p & 875.85 & O   & 45.04  & 875.85 & -\tabularnewline
\hline 
G\_100x100x2\_a1474-p & 1057.83 & O   & 46.75  & 1057.83 & -\tabularnewline
\hline 
G\_100x100x2\_a1947-p & 1234.77 & O   & 43.86  & 1234.77 & -\tabularnewline
\hline 
G\_100x100x2\_a2421-p & 1395.98 & O   & 46.19  & 1395.98 & -\tabularnewline
\hline 
G\_100x100x2\_a2895-p & 1554.88 & O   & 47.62  & 1554.88 & -\tabularnewline
\hline 
G\_100x100x2\_a3368-p & 1697.28 & O   & 46.85  & 1697.28 & -\tabularnewline
\hline 
G\_100x100x2\_a3842-p & 1824.59 & N   & 34.68  & 1833.18 & -0.47\tabularnewline
\hline 
G\_100x100x2\_a4316-p & 1941.98 & N   & 53.69  & 1955.73 & -0.70\tabularnewline
\hline 
G\_100x100x2\_a4789-p & 2036.87 & NF   & 241.08  & 2056.08 & -0.93\tabularnewline
\hline 
G\_100x100x2\_a5263-p & 2122.72 & NF   & 230.64  & 2152.41 & -1.38\tabularnewline
\hline 
G\_100x100x2\_a5737-p & 2229.83 & NF   & 169.36  & 2264.99 & -1.55\tabularnewline
\hline 
G\_100x100x2\_a6211-p & 2339.31 & N   & 40.47 & 2382.05 & -1.79\tabularnewline
\hline 
G\_100x100x2\_a6684-p & 2414.71 & N   & 516.42  & 2457.86 & -1.76\tabularnewline
\hline 
G\_100x100x2\_a7158-p & 2524.14 & N   & 52.56  & 2575.23 & -1.98\tabularnewline
\hline 
G\_100x100x2\_a7632-p & 2609.18 & NF   & 243.35  & 2658.92 & -1.87\tabularnewline
\hline 
G\_100x100x2\_a8105-p & 2700.71 & NF   & 456.12  & 2758.35 & -2.09\tabularnewline
\hline 
G\_100x100x2\_a8579-p & 2775.56 & N   & 34.92  & 2834.97 & -2.10\tabularnewline
\hline 
G\_100x100x2\_a9053-p & 2863.76 & NF   & 362.59  & 2919.05 & -1.89\tabularnewline
\hline 
G\_100x100x2\_a9526-p & 2953.47 & N   & 211.22  & 3019.45 & -2.19\tabularnewline
\hline 
G\_100x100x2\_a10000-p & 3012.83 & N   & 42.95  & 3074.09 & -1.99\tabularnewline
\end{tabular}
\par\end{centering}

\caption{Primal Bounds for the PCSPG on 3d grid graphs with many terminals\label{tab:G3DPCmany}}

\end{table}

\pagebreak{}

\section{Dimacs and Stenlib results \label{sec:Dimacs=000026StenLib}}

\begin{table}[H]
\noindent \begin{centering}
\begin{tabular}{l|llr|l.}
\multicolumn{1}{c|}{\textbf{Instance}} & \multicolumn{1}{c}{\textbf{New PB}} & \multicolumn{1}{l}{\textbf{Algo}} & \multicolumn{1}{r|}{\textbf{Time {[}s{]}}} & \multicolumn{1}{c}{\textbf{``polito'' PB}} & \multicolumn{1}{c}{\textbf{Gap \%}}\tabularnewline
\hline 
d18 & 223 & All & 2.97 (Fpre) & 224 & -0.45\tabularnewline
\hline 
e18 & 564 & All & 112.78 (NFpre) & 565 & -0.18\tabularnewline
\hline 
i640-211 & 11984 & All & 851.79 (Opre) & 11984 & 0.00\tabularnewline
\hline 
i640-314 & 35532 & N NF & 223.53 (N) & 35532 & 0.00\tabularnewline
\hline 
i640-341 & 32042 & NFpre & 2164.31 & 32047 & -0.02\tabularnewline
\hline 
fnl4461fst & 189367 & Npre & 3759.34 & 192847 & -1.80\tabularnewline
\hline 
world666 & 122971 & N & 94.37 & 130516 & -5.78\tabularnewline
\hline 
alue7080 & 66624 & NF & 100.96 & 67847 & -1.80\tabularnewline
\hline 
alut2625 & 40183 & Npre & 1042.46 & 41501 & -3.18\tabularnewline
\hline 
es10000fst01 & 733237957 & Npre & 1232.63 & 764631264 & -4.11\tabularnewline
\hline 
lin36 & 64486 & Npre & 399.04 & 64052 & 0.68\tabularnewline
\hline 
lin37 & 112886 & Npre & 892.89 & 114001 & -0.98\tabularnewline
\hline 
hc12p & 236075 & N Npre & 3664.50 (N) & 236042 & 0.01\tabularnewline
\hline 
hc12u & 2269 & Opre J Jpre N Npre & 4883.88 (Npre) & 2265 & 0.18\tabularnewline
\hline 
cc12-2p & 121056 & N & 7077.52 & 121091 & -0.03\tabularnewline
\hline 
cc12-2u & 1174 & N Npre & 6330.21 (Npre) & 1177 & -0.25\tabularnewline
\hline 
cc12-2n & 613 & Npre & 4425.05 & 615 & -0.33\tabularnewline
\hline 
cc3-12n & 111 & All & 20.60 (Opre) & 111 & 0.00\tabularnewline
\hline 
cc3-12p & 19003 & N Npre & 5750.94 (N) & 18932 & 0.38\tabularnewline
\hline 
cc3-12u & 185 & All & 2331.35 (N) & 186 & -0.54\tabularnewline
\hline 
bipa2p & 35285 & Npre & 1927.36 & 35336 & -0.14\tabularnewline
\hline 
bipa2u & 338 & All & 118.86 (NF) & 339 & -0.29\tabularnewline
\hline 
2r211c & 89000 & Opre J Jpre N Npre & 615.75 (Jpre) & 90000 & -1.11\tabularnewline
\hline 
wrp3-83 & 8301057 & N & 1191.62 & 8301263 & 0.00\tabularnewline
\hline 
w23c23 & 691 & F Fpre JF JFpre NF NFpre & 3541.60 (NF) & 693 & -0.29\tabularnewline
\hline 
rc09 & 120450 & Npre & 4176.39 & 122358 & -1.56\tabularnewline
\hline 
rt05 & 58286 & F & 27.08 (F) & 59147 & -1.46\tabularnewline
\hline 
G106ac & 40396063 & N  & 7088.62 & 40346932 & 0.12\tabularnewline
\hline 
I064ac & 188849475 & J & 4678.88 & 188479558 & 0.20\tabularnewline
\hline 
s5 & 25210 & All & 124.29 (NF) & 25210 & 0.00\tabularnewline
\hline 
\end{tabular}
\par\end{centering}

\caption{Best PB for all the SPG instances and comparison with ``polito''
results}

\label{PB_SPG_table}
\end{table}

\pagebreak{}

\begin{table}[H]
\begin{centering}
\begin{tabular}{l|lr.}
\multicolumn{1}{c|}{\textbf{Instance}} & \multicolumn{1}{l}{\textbf{PB}} & \multicolumn{1}{r}{\textbf{Best PB of the Challenge}} & \multicolumn{1}{c}{\textbf{Gap \%}}\tabularnewline
\hline 
d18 & 223 & 223 (AB mozartballs scipjack staynerd) & 0.00\tabularnewline
\hline 
e18 & 564 & 564 (mozartballs scipjack staynerd) & 0.00\tabularnewline
\hline 
i640-211 & 11984 & 11984 (polito scipjack) & 0.00\tabularnewline
\hline 
i640-314 & 35532 & 35532 (polito staynerd) & 0.00\tabularnewline
\hline 
i640-341 & \textbf{32042} & 32047 (polito) & -0.02\tabularnewline
\hline 
fnl4461fst & 189367 & 182527 (PUW) & 3.75\tabularnewline
\hline 
world666 & 122971 & 122467 (mozartballs PUW scipjack staynerd) & 0.41\tabularnewline
\hline 
alue7080 & 66624 & 62514 (PUW) & 6.57\tabularnewline
\hline 
alut2625 & 40183 & 35471 (PUW) & 13.28\tabularnewline
\hline 
es10000fst01 & 733237957 & 716559567 (mozartballs) & 2.33\tabularnewline
\hline 
lin36 & 64486 & 55608 (PUW) & 15.97\tabularnewline
\hline 
lin37 & 112886 & 99560 (PUW) & 13.38\tabularnewline
\hline 
hc12p & 236075 & 236042 (polito) & 0.01\tabularnewline
\hline 
hc12u & 2269 & 2262 (staynerd) & 0.31\tabularnewline
\hline 
cc12-2p & \textbf{121056} & 121091 (polito) & -0.03\tabularnewline
\hline 
cc12-2u & \textbf{1174} & 1177 (polito) & -0.25\tabularnewline
\hline 
cc12-2n & \textbf{613} & 615 (polito) & -0.33\tabularnewline
\hline 
cc3-12n & 111 & 111 (mozartballs polito PUW scipjack staynerd) & 0.00\tabularnewline
\hline 
cc3-12p & 19003 & 18865 (mozartballs) & 0.73\tabularnewline
\hline 
cc3-12u & 185 & 185 (mozartballs PUW staynerd) & 0.00\tabularnewline
\hline 
bipa2p & \textbf{35285} & 35336 (polito) & -0.14\tabularnewline
\hline 
bipa2u & 338 & 337 (mozartballs staynerd) & 0.30\tabularnewline
\hline 
2r211c & 89000 & 89000 (mozartballs PUW scipjack staynerd) & 0.00\tabularnewline
\hline 
wrp3-83 & 8301057 & 8300906 (PUW) & 0.00\tabularnewline
\hline 
w23c23 & 691 & 689 (mozartballs staynerd) & 0.29\tabularnewline
\hline 
rc09 & 120450 & 111005 (mozartballs staynerd) & 8.51\tabularnewline
\hline 
rt05 & 58286 & 51354 (PUW) & 13.50\tabularnewline
\hline 
G106ac & 40396063 & 36920936 (mozartballs staynerd) & 9.41\tabularnewline
\hline 
I064ac & 188849475 & 186852309 (PUW) & 1.07\tabularnewline
\hline 
s5 & 25210 & 25210 (mozartballs polito PUW scipjack staynerd) & 0.00\tabularnewline
\hline 
\end{tabular}
\par\end{centering}

\caption{Best PB for all the SPG instances and comparison with the best results
of the Challenge}

\label{PB_other}
\end{table}

\pagebreak{}

\begin{table}[H]
\begin{centering}
\begin{tabular}{l|lcr|l.}
\multicolumn{1}{c|}{\textbf{Instance}} & \multicolumn{1}{l}{\textbf{PB}} & \multicolumn{1}{c}{\textbf{Algo}} & \multicolumn{1}{r|}{\textbf{Time {[}s{]}}} & \multicolumn{1}{c}{\textbf{``polito'' PB }} & \multicolumn{1}{c}{\textbf{Gap \%}}\tabularnewline
\hline 
C13-A & 236 & F J N W & 1.23 (N) & 237 & -0.42 \tabularnewline
\hline 
C19-B & 146 & F J N W & 2.62 (N) & 146 & 0.00\tabularnewline
\hline 
D03-B & 1509 & F J N W & 5.79 (N) & 1510 & -0.07\tabularnewline
\hline 
D20-A & 536 & F J N W & 4.64 (N) & 536 & 0.00\tabularnewline
\hline 
P400-3 & 2951725 & F J N W & 1.64 (N) & 2951725 & 0.00\tabularnewline
\hline 
P400-4 & 2852956 & F J N W & 4.30 (N) & 2852956 & 0.00\tabularnewline
\hline 
K400-7 & 485587 & W & 2289.60 & 523885 & -7.31\tabularnewline
\hline 
K400-10 & 401032 & W & 1086.08 & 406365 & -1.31\tabularnewline
\hline 
hc10p & 59682 & W & 7140.50 & 59813 & -0.22\tabularnewline
\hline 
hc11u & 1115 & F J N & 1988.85 (N) & 1120 & -0.45\tabularnewline
\hline 
hc12p & 235132 & F J N & 5463.03 (N) & 235043 & 0.04\tabularnewline
\hline 
hc12u & 2216 & F J N & 6071.29 (N) & 2227 & -0.49\tabularnewline
\hline 
bip52nu & 222 & F & 3093.53 & 223 & -0.45\tabularnewline
\hline 
bip62nu & 214 & F J N W & 4.32 (N) & 214 & 0.00\tabularnewline
\hline 
cc3-12nu & 95 & F J N W & 26.43 (N) & 95 & 0.00\tabularnewline
\hline 
cc12-2nu & 565 & F N & 2789.17 (F) & 567 & -0.35\tabularnewline
\hline 
i640-001 & 2932 & F J N W & 0.42 (N) & 3053 & -3.96\tabularnewline
\hline 
i640-221 & 8430 & F & 5466.72 & 8626 & -2.27\tabularnewline
\hline 
i640-321 & 28790 & F & 6949.88 & 28821 & -0.11\tabularnewline
\hline 
i640-341 & 29679 & F & 2037.28 & 29713 & -0.11\tabularnewline
\hline 
a2000RandGraph-2 & 1483.84 & F J N W & 10.04 (N) & 1484.2 & -0.02\tabularnewline
\hline 
a4000RandGraph-3 & 3406.62 & F J N W & 6.74 (N) & 3407.5 & -0.03\tabularnewline
\hline 
a8000RandGraph-1-2 & 4719.97 & F J N & 5552.40 (N) & 4722.8 & -0.06\tabularnewline
\hline 
a14000RandGraph-1-5 & 9475.59 & F J N & 690.66 (N) & 9475.6 & 0.00\tabularnewline
\hline 
handsd04 & 525.86 & W & 1470.15 & 584.1 & -9.97\tabularnewline
\hline 
handbd13 & 13.23 & F J N W & 100.15 (N) & 18.1 & -26.91\tabularnewline
\hline 
handsi03 & 56.23 & F J N & 31.95 (J) & 56.3 & -0.12\tabularnewline
\hline 
handbi07 & 151.04 & F J N W & 88.46 (W) & 151.1 & -0.04\tabularnewline
\hline 
drosophila001 & 8273.98 & W & 3491.73 & 8288.3 & -0.17\tabularnewline
\hline 
HCMV & 7376.36 & F J N W & 4.62 (W) & 7378.2 & -0.02\tabularnewline
\hline 
lymphoma & 3341.89 & F J N W & 344.67 (N) & 3349.1 & -0.22\tabularnewline
\hline 
metabol-expr-mice-1 & 11346.93 & F J N W & 2120.98 (W) & 11901.9 & -4.66\tabularnewline
\hline 
\end{tabular}
\par\end{centering}

\caption{Best PB for all the PCSPG instances and comparison with ``polito''
results}

\label{PB_PC_best}
\end{table}

\pagebreak{}

\begin{table}[H]
\begin{centering}
\begin{tabular}{l|lr.}
\multicolumn{1}{c|}{\textbf{Instance}} & \multicolumn{1}{l}{\textbf{PB}} & \multicolumn{1}{r}{\textbf{Best PB of the Challenge}} & \multicolumn{1}{c}{\textbf{Gap \%}}\tabularnewline
\hline 
C13-A & 236 & 236 (KTS mozartballs staynerd) & 0.00\tabularnewline
\hline 
C19-B & 146 & 146 (All) & 0.00\tabularnewline
\hline 
D03-B & 1509 & 1509 (All) & 0.00\tabularnewline
\hline 
D20-A & 536 & 536 (All) & 0.00\tabularnewline
\hline 
P400-3 & 2951725 & 2951725 (All) & 0.00\tabularnewline
\hline 
P400-4 & 2852956 & 2852956 (All) & 0.00\tabularnewline
\hline 
K400-7 & 485587 & 474466 (KTS mozartballs staynerd) & 2.34\tabularnewline
\hline 
K400-10 & 401032 & 394191 (KTS mozartballs staynerd) & 1.74\tabularnewline
\hline 
hc10p & \textbf{59682} & 59738 (polito) & -0.09\tabularnewline
\hline 
hc11u & \textbf{1115} & 1116 (mozartballs staynerd) & -0.09\tabularnewline
\hline 
hc12p & 235132 & 234977 (polito) & 0.07\tabularnewline
\hline 
hc12u & \textbf{2216} & 2221 (KTS) & -0.23\tabularnewline
\hline 
bip52nu & 222 & 222 (mozartballs) & 0.00\tabularnewline
\hline 
bip62nu & 214 & 214 (All) & 0.00\tabularnewline
\hline 
cc3-12nu & 95 & 95 (KTS mozartballs staynerd) & 0.00\tabularnewline
\hline 
cc12-2nu & \textbf{565} & 567 (KTS polito) & -0.35\tabularnewline
\hline 
i640-001 & 2932 & 2932 (All) & 0.00\tabularnewline
\hline 
i640-221 & 8430 & 8400 (KTS mozartballs) & 0.36\tabularnewline
\hline 
i640-321 & 28790 & 28787 (KTS) & 0.01\tabularnewline
\hline 
i640-341 & 29679 & 29666 (KTS) & 0.04\tabularnewline
\hline 
a2000RandGraph-2 & 1483.84 & 1483.8 (mozartballs staynerd) & 0.00\tabularnewline
\hline 
a4000RandGraph-3 & 3406.62 & 3406.6 (mozartballs staynerd) & 0.00\tabularnewline
\hline 
a8000RandGraph-1-2 & 4719.97 & 4720.0 (mozartballs staynerd) & 0.00\tabularnewline
\hline 
a14000RandGraph-1-5 & 9475.59 & 9475.6 (mozartballs staynerd) & 0.00\tabularnewline
\hline 
handsd04 & 525.86 & 493.80 (staynerd) & 6.49\tabularnewline
\hline 
handbd13 & 13.23 & 13.20 (All) & 0.23\tabularnewline
\hline 
handsi03 & 56.23 & 56.10 (mozartballs staynerd) & 0.23\tabularnewline
\hline 
handbi07 & 151.04 & 151 (KTS mozartballs staynerd) & 0.03\tabularnewline
\hline 
drosophila001 & 8273.98 & 8274 (mozartballs staynerd) & 0.00\tabularnewline
\hline 
HCMV & 7376.36 & 7371.5 (KTS mozartballs staynerd) & 0.07\tabularnewline
\hline 
lymphoma & 3341.89 & 3341.9 (KTS mozartballs staynerd) & 0.00\tabularnewline
\hline 
metabol-expr-mice-1 & 11346.93 & 11346.9 (mozartballs staynerd) & 0.00\tabularnewline
\hline 
\end{tabular}
\par\end{centering}

\caption{Best PB for all the PCSPG instances and comparison with the best results
of the Challenge}

\label{PB_PC_chal}
\end{table}

\pagebreak{}

\begin{table}[H]
\noindent \begin{centering}
\begin{tabular}{l|lrr}
\multicolumn{1}{c|}{\textbf{Instance}} & \multicolumn{1}{c}{\textbf{MS reweighting}} & \multicolumn{1}{r}{\textbf{MST}} & \multicolumn{1}{r}{\textbf{SPT}}\tabularnewline
\hline 
d18 & 223 & 335 & 358\tabularnewline
\hline 
e18 & 564 & 902 & 892\tabularnewline
\hline 
i640-211 & 11984 & 26928 & 15967\tabularnewline
\hline 
i640-314 & 35532 & 55479 & 41186\tabularnewline
\hline 
i640-341 & 32042 & 53826 & 43593\tabularnewline
\hline 
fnl4461fst & 189367 & 203167 & 271681\tabularnewline
\hline 
world666 & 122971 & 172983 & 1356820\tabularnewline
\hline 
alue7080 & 66624 & 112478 & 188092\tabularnewline
\hline 
alut2625 & 40183 & 89813 & 157660\tabularnewline
\hline 
es10000fst01 & 733237957 & 760866530 & 990978452\tabularnewline
\hline 
lin36 & 64486 & 145363 & 130588\tabularnewline
\hline 
lin37 & 112886 & 223493 & 233031\tabularnewline
\hline 
hc12p & 236075 & 332813 & 314607\tabularnewline
\hline 
hc12u & 2269 & 2749 & 2878\tabularnewline
\hline 
cc12-2p & 121056 & 237979 & 175225\tabularnewline
\hline 
cc12-2u & 1174 & 1644 & 1624\tabularnewline
\hline 
cc12-2n & 613 & 1130 & 1116\tabularnewline
\hline 
cc3-12n & 111 & 165 & 155\tabularnewline
\hline 
cc3-12p & 19003 & 44366 & 26078\tabularnewline
\hline 
cc3-12u & 185 & 245 & 255\tabularnewline
\hline 
bipa2p & 35285 & 56488 & 45663\tabularnewline
\hline 
bipa2u & 338 & 421 & 420\tabularnewline
\hline 
2r211c & 89000 & 155000 & 165000\tabularnewline
\hline 
wrp3-83 & 8301057 & 8301505 & 8402368\tabularnewline
\hline 
w23c23 & 691 & 828 & 918\tabularnewline
\hline 
rc09 & 120450 & 138800 & 226297\tabularnewline
\hline 
rt05 & 58286 & 66236 & 103866\tabularnewline
\hline 
G106ac & 40396063 & 44792419 & 45158294\tabularnewline
\hline 
I064ac & 188849475 & 190392002 & 192238839\tabularnewline
\hline 
s5 & 25210 & 25210 & 25210\tabularnewline
\hline 
\end{tabular}
\par\end{centering}

\caption{Heuristics vs MS guided heuristics. Comparison of the PB for the SPG
instances}

\label{Heur_MS}
\end{table}

\pagebreak{}

\begin{table}[H]
\noindent \begin{centering}
\begin{tabular}{l|lrrr}
\multicolumn{1}{c|}{\textbf{Instance}} & \multicolumn{1}{c}{\textbf{MS reweightng}} & \multicolumn{1}{r}{\textbf{MST}} & \multicolumn{1}{r}{\textbf{SPT}} & \multicolumn{1}{r}{\textbf{GW}}\tabularnewline
\hline 
C13-A & 236 & 305 & 360 & 309\tabularnewline
\hline 
C19-B & 146 & 215 & 237 & 204\tabularnewline
\hline 
D03-B & 1509 & 1819 & 1799 & 1819\tabularnewline
\hline 
D20-A & 536 & 661 & 638 & 806\tabularnewline
\hline 
P400-3 & 2951725 & 3633310 & 4108109 & 3475372\tabularnewline
\hline 
P400-4 & 2852956 & 3649056 & 3785708 & 3577568\tabularnewline
\hline 
K400-7 & 485587 & 528817 & 528817 & 528817\tabularnewline
\hline 
K400-10 & 401032 & 434488 & 434488 & 434488\tabularnewline
\hline 
hc10p & 59682 & 77232 & 77232 & 75181\tabularnewline
\hline 
hc11u & 1115 & 1318 & 1363 & 1264\tabularnewline
\hline 
hc12p & 235132 & 308065 & 308065 & 308065\tabularnewline
\hline 
hc12u & 2216 & 2581 & 2666 & 2543\tabularnewline
\hline 
bip52nu & 222 & 284 & 281 & 278\tabularnewline
\hline 
bip62nu & 214 & 239 & 242 & 244\tabularnewline
\hline 
cc3-12nu & 95 & 112 & 112 & 112\tabularnewline
\hline 
cc12-2nu & 565 & 697 & 697 & 697\tabularnewline
\hline 
i640-001 & 2932 & 3764 & 3764 & 3764\tabularnewline
\hline 
i640-221 & 8430 & 21117 & 12006 & 11630\tabularnewline
\hline 
i640-321 & 28790 & 49605 & 42401 & 41880\tabularnewline
\hline 
i640-341 & 29679 & 49707 & 38085 & 41826\tabularnewline
\hline 
a2000RandGraph-2 & 1483.84 & 1648.58 & 1822.47 & 1645.15\tabularnewline
\hline 
a4000RandGraph-3 & 3406.62 & 3616.03 & 4314.37 & 3615.58\tabularnewline
\hline 
a8000RandGraph-1-2 & 4719.97 & 4790.86 & 4790.86 & 4790.86\tabularnewline
\hline 
a14000RandGraph-1-5 & 9475.59 & 10514.12 & 10514.12 & 10514.12\tabularnewline
\hline 
handsd04 & 525.86 & 792.61 & 792.61 & 792.61\tabularnewline
\hline 
handbd13 & 13.23 & 13.24 & 13.24 & 13.24\tabularnewline
\hline 
handsi03 & 56.23 & 56.26 & 56.26 & 56.26\tabularnewline
\hline 
handbi07 & 151.04 & 151.06 & 151.06 & 151.06\tabularnewline
\hline 
drosophila001 & 8273.98 & 8296.30 & 8296.30 & 8296.30\tabularnewline
\hline 
HCMV & 7376.36 & 7376.84 & 7376.84 & 7376.84\tabularnewline
\hline 
lymphoma & 3341.89 & 3410.36 & 3410.36 & 3410.36\tabularnewline
\hline 
metabol-expr-mice-1 & 11346.93 & 11885.90 & 11885.90 & 11885.90\tabularnewline
\hline 
\end{tabular}
\par\end{centering}

\caption{Heuristics vs MS guided heuristics. Comparison of the PB for the PCSPG
instances}

\label{Heur_MS2}
\end{table}

\pagebreak{}

\begin{table}[H]
\begin{centering}
\begin{tabular}{l|lr.}
\multicolumn{1}{c|}{\textbf{Instance}} & \multicolumn{1}{c}{\textbf{Best Known}} & \multicolumn{1}{r}{\textbf{Npre}} & \multicolumn{1}{c}{\textbf{Gap \%}}\tabularnewline
\hline 
bip42p & 24657 (opt) & 24657 & 0.00\tabularnewline
\hline 
bip42u & 236 (opt) & 236 & 0.00\tabularnewline
\hline 
bip52p & 24526 & 24549 & 0.09\tabularnewline
\hline 
bip52u & 234 & 234 & 0.00\tabularnewline
\hline 
bip62p & 22843 & 22843 & 0.00\tabularnewline
\hline 
bip62u & 219 & 219 & 0.00\tabularnewline
\hline 
bipe2p & 5616 (opt) & 5616 & 0.00\tabularnewline
\hline 
bipe2u & 54 (opt) & 54 & 0.00\tabularnewline
\hline 
cc10-2p & 35297 & \textbf{35269} & -0.08\tabularnewline
\hline 
cc10-2u & 342 & 342 & 0.00\tabularnewline
\hline 
cc11-2p & 63491 & \textbf{63405} & -0.14\tabularnewline
\hline 
cc11-2u & 612 & 614 & 0.33\tabularnewline
\hline 
cc3-10p & 12772 & 12870 & 0.77\tabularnewline
\hline 
cc3-10u & 125 & 125 & 0.00\tabularnewline
\hline 
cc3-11p & 15582 & 15680 & 0.63\tabularnewline
\hline 
cc3-11u & 153 & 153 & 0.00\tabularnewline
\hline 
cc3-4p & 2338 (opt) & 2338 & 0.00\tabularnewline
\hline 
cc3-4u & 23 (opt) & 23 & 0.00\tabularnewline
\hline 
cc3-5p & 3661 (opt) & 3665 & 0.11\tabularnewline
\hline 
cc3-5u & 36 (opt) & 36 & 0.00\tabularnewline
\hline 
cc5-3p & 7299 (opt) & 7302 & 0.04\tabularnewline
\hline 
cc5-3u & 71 (opt) & 71 & 0.00\tabularnewline
\hline 
cc6-2p & 3271 (opt) & 3271 & 0.00\tabularnewline
\hline 
cc6-2u & 32 (opt) & 32 & 0.00\tabularnewline
\hline 
cc6-3p  & 20270 (opt) & 20298 & 0.14\tabularnewline
\hline 
cc6-3u & 197 (opt) & 198 & 0.51\tabularnewline
\hline 
cc7-3p & 56799 & 56835 & 0.06\tabularnewline
\hline 
cc7-3u & 549 & 553 & 0.73\tabularnewline
\hline 
cc9-2p & 17199 & 17225 & 0.15\tabularnewline
\hline 
cc9-2u & 167 (opt) & 167 & 0.00\tabularnewline
\hline 
\end{tabular}
\par\end{centering}

\caption{Best known energies for PUC instances, ``\textit{bip}''-like and
``\textit{cc}''-like classes }

\label{PUC}
\end{table}

\pagebreak{}

\begin{table}[H]
\begin{centering}
\begin{tabular}{l|lr.}
\multicolumn{1}{c|}{\textbf{Instance}} & \multicolumn{1}{c}{\textbf{Best Known}} & \multicolumn{1}{r}{\textbf{Npre}} & \multicolumn{1}{c}{\textbf{Gap \%}}\tabularnewline
\hline 
hc10p & 59797 & 59808 & 0.02\tabularnewline
\hline 
hc10u & 575 & 575 & 0.00\tabularnewline
\hline 
hc11p & 119492 & \textbf{119456} & -0.03\tabularnewline
\hline 
hc11u & 1145 & 1153 & 0.70\tabularnewline
\hline 
hc6p & 4003 (opt) & 4003 & 0.00\tabularnewline
\hline 
hc6u & 39 (opt) & 39 & 0.00\tabularnewline
\hline 
hc7p & 7905 (opt) & 7906 & 0.01\tabularnewline
\hline 
hc7u & 77 (opt) & 77 & 0.00\tabularnewline
\hline 
hc8p & 15322 (opt) & 15336 & 0.09\tabularnewline
\hline 
hc8u & 148 (opt) & 148 & 0.00\tabularnewline
\hline 
hc9p & 30242 & 30313 & 0.23\tabularnewline
\hline 
hc9u & 292 & 292 & 0.00\tabularnewline
\hline 
\end{tabular}
\par\end{centering}

\caption{Best known energies for PUC instances, ``\textit{hc}''-like class}

\label{PUC2}

\end{table}

\pagebreak{}
%\end{document}


\begin{thebibliography}{10}

\bibitem{huang_integrating_2009}
Shao-shan~Carol Huang and Ernest Fraenkel.
\newblock Integrating {Proteomic}, {Transcriptional}, and {Interactome} {Data}
  {Reveals} {Hidden} {Components} of {Signaling} and {Regulatory} {Networks}.
\newblock {\em Sci. Signal.}, 2(81):ra40--ra40, July 2009.

\bibitem{bailly-bechet_prize-collecting_2009}
Marc Bailly-Bechet, Alfredo Braunstein, and Riccardo Zecchina.
\newblock A prize-collecting steiner tree approach for transduction network
  inference.
\newblock In Pierpaolo Degano and Roberto Gorrieri, editors, {\em Computational
  Methods in Systems Biology}, number 5688 in Lecture Notes in Computer
  Science, pages 83--95. Springer Berlin Heidelberg, January 2009.

\bibitem{tuncbag_simultaneous_2012}
Nurcan Tuncbag, Alfredo Braunstein, Andrea Pagnani, Shao-ShanCarol Huang,
  Jennifer Chayes, Christian Borgs, Riccardo Zecchina, and Ernest Fraenkel.
\newblock Simultaneous reconstruction of multiple signaling pathways via the
  prize-collecting steiner forest problem.
\newblock In Benny Chor, editor, {\em Research in Computational Molecular
  Biology}, volume 7262 of {\em Lecture Notes in Computer Science}, pages
  287--301. Springer Berlin Heidelberg, 2012.
\newblock Cited by 0002.

\bibitem{ljubic_algorithmic_2005}
Ivana Ljubi{\'c}, Ren{\'e} Weiskircher, Ulrich Pferschy, Gunnar~W. Klau, Petra
  Mutzel, and Matteo Fischetti.
\newblock An {Algorithmic} {Framework} for the {Exact} {Solution} of the
  {Prize}-{Collecting} {Steiner} {Tree} {Problem}.
\newblock {\em Math. Program.}, 105(2-3):427--449, October 2005.

\bibitem{bayati_rigorous_2008}
Mohsen Bayati, A.~Braunstein, and Riccardo Zecchina.
\newblock A rigorous analysis of the cavity equations for the minimum spanning
  tree.
\newblock {\em Journal of Mathematical Physics}, 49(12):125206, 2008.
\newblock Cited by 0012.

\bibitem{bayati_statistical_2008}
M.~Bayati, C.~Borgs, A.~Braunstein, J.~Chayes, A.~Ramezanpour, and R.~Zecchina.
\newblock Statistical mechanics of steiner trees.
\newblock {\em Physical Review Letters}, 101(3):037208, July 2008.

\bibitem{biazzo_performance_2012}
Indaco Biazzo, Alfredo Braunstein, and Riccardo Zecchina.
\newblock Performance of a cavity-method-based algorithm for the
  prize-collecting steiner tree problem on graphs.
\newblock {\em Phys. Rev. E}, 86:026706, August 2012.

\bibitem{mezard_information_2009}
Marc M{\'e}zard and Andrea Montanari.
\newblock {\em Information, Physics, and Computation}.
\newblock Oxford University Press, January 2009.

\bibitem{bailly-bechet_finding_2011}
M.~Bailly-Bechet, C.~Borgs, A.~Braunstein, J.~Chayes, A.~Dagkessamanskaia,
  J.-M. Fran{\c c}ois, and R.~Zecchina.
\newblock Finding undetected protein associations in cell signaling by belief
  propagation.
\newblock {\em Proceedings of the National Academy of Sciences},
  108(2):882--887, January 2011.

\bibitem{Goemans:1992:GAT:139404.139468}
Michel~X. Goemans and David~P. Williamson.
\newblock A general approximation technique for constrained forest problems.
\newblock In {\em Proceedings of the Third Annual ACM-SIAM Symposium on
  Discrete Algorithms}, SODA '92, pages 307--316, Philadelphia, PA, USA, 1992.
  Society for Industrial and Applied Mathematics.

\bibitem{Goemans:1996:PMA:241938.241942}
Michel~X. Goemans and David~P. Williamson.
\newblock Approximation algorithms for np-hard problems.
\newblock chapter The Primal-dual Method for Approximation Algorithms and Its
  Application to Network Design Problems, pages 144--191. PWS Publishing Co.,
  Boston, MA, USA, 1997.

\end{thebibliography}
\end{document}